\documentclass[10pt,preprint]{aastex}
\usepackage{natbib}
\usepackage{graphicx}
\usepackage{epsfig}
\usepackage{amssymb}

\newcommand{\Barnes}{{C. Barnes}}
\newcommand{\Bennett}{{C. L. Bennett}}

\newcommand{\Halpern}{{M. Halpern}}
\newcommand{\Hill}{{R. S. Hill}}
\newcommand{\Hinshaw}{{G. Hinshaw}}
\newcommand{\Jarosik}{{N. Jarosik}}
\newcommand{\Kogut}{{A. Kogut}}
\newcommand{\Komatsu}{{E. Komatsu}}
\newcommand{\Limon}{{M. Limon}}
\newcommand{\Meyer}{{S. S. Meyer}}

\newcommand{\Page}{{L. Page}}

\newcommand{\Spergel}{{D. N. Spergel}}
\newcommand{\Tucker}{{G. S. Tucker}}
\newcommand{\Verde}{{L. Verde}}
\newcommand{\Weiland}{{J. L. Weiland}}
\newcommand{\Wollack}{{E. Wollack}}
\newcommand{\Wright}{{E. L. Wright}}

\newcommand{\Brown}{{Dept. of Physics, Brown University, 
            Providence, RI 02912}}
\newcommand{\Goddard}{{Code 685, Goddard Space Flight Center, 
            Greenbelt, MD 20771}}
\newcommand{\NRCFellow}{{National Research Council (NRC) Fellow}}
\newcommand{\PrincetonPhysics}{{Dept. of Physics, Jadwin Hall, 
            Princeton, NJ 08544}}
\newcommand{\PrincetonAstro}{{Dept of Astrophysical Sciences, 
            Princeton University, Princeton, NJ 08544}}
\newcommand{\SSAI}{{Science Systems and Applications, Inc. (SSAI), 
            10210 Greenbelt Road, Suite 600 Lanham, Maryland 20706}}
\newcommand{\UBC}{{Dept. of Physics and Astronomy, University of 
            British Columbia, Vancouver, BC  Canada V6T 1Z1}}
\newcommand{\UChicago}{{Depts. of Astrophysics and Physics, EFI and CfCP, 
            University of Chicago, Chicago, IL 60637}}
\newcommand{\UCLA}{{UCLA Astronomy, PO Box 951562, Los Angeles, CA 90095-1562}}

\shorttitle{{\sl WMAP} Power Spectrum}
\shortauthors{Hinshaw et al.}

\newcommand{\map}          {{\sl WMAP}}
\newcommand{\cobe}         {{\sl COBE}}
\newcommand{\n}            {{\bf{n}}}
\newcommand{\VEV}[1]       {\langle#1\rangle}
\newcommand{\lmax}         {l_{\rm max}}
\newcommand{\dg}           {\mbox{$^{\circ}$}}

\newcommand{\lsim}         {\mbox{$_<\atop^{\sim}$}}
\newcommand{\gsim}         {\mbox{$_>\atop^{\sim}$}}

\newcommand{\amin}         {\mbox{$^\prime\ $}}
\newcommand{\asec}         {\mbox{$^{\prime\prime}\ $}}
\newcommand{\ddeg}         {\mbox{${\rlap.}^\circ$}}

\newcommand{\wjjj}[6]
{{\left( \begin{array}{lcr} #1 & #2 & #3 \\#4 & #5 & #6 \end{array}\right) }}

\slugcomment{Submitted to the Astrophysical Journal}

\begin{document}

\title{First Year {\sl Wilkinson Microwave Anisotropy Probe (WMAP\altaffilmark{1})} Observations:\\
The Angular Power Spectrum}

\author{\Hinshaw \altaffilmark{2},
\Spergel \altaffilmark{3},
\Verde \altaffilmark{3},
\Hill \altaffilmark{4},
\Meyer \altaffilmark{5},
\Barnes \altaffilmark{6},
\Bennett \altaffilmark{2},
\Halpern \altaffilmark{7},
\Jarosik \altaffilmark{6},
\Kogut \altaffilmark{2}, 
\Komatsu \altaffilmark{3},
\Limon \altaffilmark{2,8}, 
\Page \altaffilmark{6},
\Tucker \altaffilmark{2,8,9},
\Weiland \altaffilmark{4},
\Wollack \altaffilmark{2},
\Wright \altaffilmark{10}}

\altaffiltext{1}{\map\ is the result of a partnership between Princeton 
                 University and NASA's Goddard Space Flight Center. Scientific 
		 guidance is provided by the \map\ Science Team.}
\altaffiltext{2}{\Goddard}
\altaffiltext{3}{\PrincetonAstro}
\altaffiltext{4}{\SSAI}
\altaffiltext{5}{\UChicago}
\altaffiltext{6}{\PrincetonPhysics}
\altaffiltext{7}{\UBC}
\altaffiltext{8}{\NRCFellow}
\altaffiltext{9}{\Brown}
\altaffiltext{10}{\UCLA}

\email{Gary.F.Hinshaw@nasa.gov}

\begin{abstract}
We present the angular power spectrum derived from the first-year {\sl
Wilkinson Microwave Anisotropy Probe (WMAP)} sky maps. We study a variety of
power spectrum estimation methods and data combinations and demonstrate that
the results are robust.  The data are  modestly contaminated by diffuse
Galactic foreground emission, but we show that a simple Galactic template model
is sufficient to remove the signal. Point sources produce a modest
contamination in the low frequency data. After masking $\sim$700 known bright
sources from the maps, we estimate residual sources contribute $\sim$3500 
$\mu$K$^2$ at 41 GHz, and $\sim$130 $\mu$K$^2$ at 94 GHz, to the power spectrum
[$l(l+1)C_l/2\pi$] at $l=1000$. {\em Systematic errors are negligible compared
to the (modest) level of foreground emission.} Our best estimate of the power
spectrum is derived from 28 cross-power spectra of statistically independent
channels. The final spectrum is essentially independent of the noise properties
of an individual radiometer. The resulting spectrum provides a definitive
measurement of the CMB power spectrum, with uncertainties limited by cosmic
variance, up to $l \sim 350$.  The spectrum clearly exhibits a first acoustic
peak at $l=220$ and a second acoustic  peak at $l \sim 540$
\citep{page/etal:2003c}, and it provides strong support for adiabatic initial
conditions \citep{spergel/etal:2003}.  \citet{kogut/etal:2003} analyze the
$C_l^{TE}$  power spectrum, and present evidence for a relatively high optical
depth, and an early period of cosmic reionization. Among other things, this
implies that the temperature power spectrum has been suppressed by $\sim$30\%
on degree angular scales, due to secondary scattering. 
\end{abstract}


\section{INTRODUCTION}
\label{intro}

The {\sl Wilkinson Microwave Anisotropy Probe (WMAP)} mission was designed to
measure the  CMB anisotropy with unprecedented precision and accuracy on
angular scales  from the full sky to several arc minutes by producing
maps at five  frequencies from 23 to 94 GHz. The \map\  satellite mission 
\citep{bennett/etal:2003} employs a matched pair of 1.4m telescopes 
\citep{page/etal:2003} observing two areas on the sky separated by 
$\sim$141\dg. A differential radiometer \citep{jarosik/etal:2003}  with a total
of 10 feeds for each of the two sets of optics \citep{barnes/etal:2003,
page/etal:2003} measures the difference in sky brightness between the two sky
pixels. The satellite is deployed at the Earth-Sun Lagrange point, $L_2$, and 
observes the sky with a compound spin and precession that covers the full sky
every six months.  The differential data are processed on the ground to produce
full sky maps of the CMB anisotropy \citep{hinshaw/etal:2003}.

Full sky maps provide the smallest record of the CMB anisotropy without  loss
of information.  They permit a wide variety of statistics to be computed from
the data -- one of the most fundamental is the angular power spectrum of the
CMB. Indeed, if the  temperature fluctuations are Gaussian, with random phase,
then the angular  power spectrum provides a {\em complete} description of the
statistical properties of the CMB. \citet{komatsu/etal:2003} have analyzed the
first-year \map\ sky  maps to search for evidence of non-Gaussianity and find
none, aside from a  modest level of point source contamination which we account
for in this paper. Thus, the measured power spectrum may be compared to 
predictions of cosmological  models to develop constraints on model
parameters. 

This paper presents the angular power spectrum obtained from the first-year
\map\ sky maps. Companion papers present the maps and an overview of the  basic
results \citep{bennett/etal:2003b}, and describe the foreground removal process
that precedes the power spectrum analysis \citep{bennett/etal:2003c}.
\citet{spergel/etal:2003}, \citet{verde/etal:2003}, \citet{peiris/etal:2003}, 
\citet{kogut/etal:2003}, and \citet{page/etal:2003c} discuss the implications
of the \map\ power spectrum for cosmological parameters and carry out a
joint analysis of the \map\ spectrum together with other CMB data and data from
large-scale structure probes. \citet{hinshaw/etal:2003},
\citet{jarosik/etal:2003b}, \citet{page/etal:2003b}, \citet{barnes/etal:2003},
and \citet{limon/etal:2003} discuss the data processing, the radiometer
performance, the instrument beam characteristics and the spacecraft in-orbit
performance, respectively.

A sky map $\Delta T(\n)$ defined over the full sky can be decomposed in 
spherical harmonics
\begin{equation}
\Delta T(\n) = \sum_{l>0}\sum_{m=-l}^{l} a_{l m} Y_{l m}(\n)
\end{equation}
with
\begin{equation}
a_{l m}= \int d\Omega_{\n} \, \Delta T(\n) Y^*_{l m}(\n),
\label{eq:int_alm_full}
\end{equation}
where ${\bf n}$ is a unit direction vector, and $Y_{lm}({\bf n})$ are the 
spherical harmonic functions evaluated in the direction ${\bf n}$.

If the CMB temperature fluctuation $\Delta T$ is Gaussian distributed, then
each $a_{l m}$ is an independent Gaussian deviate with
\begin{equation}
\VEV{a_{l m}} = 0,
\end{equation}
and
\begin{equation}
\VEV{a_{l m} a^*_{l'm'}} = \delta_{ll'}\,\delta_{mm'}\,C_l,
\label{eq:ortho_alm}
\end{equation}
where $C_l$ is the ensemble average power spectrum predicted by models,
and $\delta$ is the Kronecker symbol. The actual power spectrum realized in 
our sky is
\begin{equation}
C^{\rm sky}_l 
= \frac{1}{2l+1}\sum_{m=-l}^{l}\left|{{a}_{l m}}\right|^2.
\label{eq:sum_alm_full}
\end{equation}
In the absence of noise, and with full sky coverage, the right hand side of
equation~(\ref{eq:sum_alm_full}) provides an unbiased estimate of the
underlying theoretical power spectrum, which is limited only by cosmic
variance. However, realistic CMB anisotropy measurements contain noise and
other sources of error that cause the quadratic estimator  in
equation~(\ref{eq:sum_alm_full}) to be biased.  In addition, while  \map\
measures the  anisotropy over the full sky, the data near the Galactic plane
are sufficiently  contaminated by foreground emission that only a portion of
the sky ($\sim$85\%) can be used for CMB power spectrum estimation.  Thus, the
integral in  equation~(\ref{eq:int_alm_full}) cannot be evaluated as such, and 
other methods must be  found to estimate $C_l$.

In Appendix~\ref{sec:ps_estimation}, we review two methods that have appeared
in the literature for estimating the angular power spectrum in the presence of
instrument noise and sky cuts.  The first \citep{hivon/etal:2002} is a
quadratic estimator that evaluates equation~(\ref{eq:int_alm_full}) on the 
{\em cut} sky yielding a ``pseudo power spectrum'' $\tilde C_l$.  The ensemble
average of this quantity is related to the true power spectrum, $C_l$ by means
of a mode coupling matrix $G_{ll'}$ \citep{hauser/peebles:1973}. The second 
method \citep{oh/spergel/hinshaw:1999} uses a maximum likelihood approach
optimized for fast evaluation with \map-like data.  In
Appendix~\ref{sec:ps_estimation} we demonstrate that the two methods produce
consistent results.

A quadratic estimator offers the possibility of computing both an
``auto-power'' spectrum, proportional to $\sum_m \vert a_{lm}\vert^2$, and a
``cross-power'' spectrum, proportional to $\sum_m a^i_{lm}a^{j*}_{lm}$, where
the $a_{lm}$ coefficients are estimated from two independent CMB maps, $i$ and
$j$. This latter form has the advantage that, if the noise in the two maps is 
uncorrelated, the quadratic estimator is not biased by the noise.   For all
cosmological analyses, we use only the cross-power spectra between
statistically independent channels. As a result, the angular spectra are, for
all intents and purposes, independent of the noise properties of an individual
radiometer. This is analogous to interferometric data which exhibits a high 
degree of immunity to systematic errors. The precise form of the estimator we
use is given in Appendix~\ref{sec:ps_estimation}.

The plan of this paper is as follows.  In \S\ref{sec:instrument} we review the
properties of the \map\ instrument and how they affect the  derived power
spectrum. In \S\ref{sec:data} we present results for the angular power spectra
obtained from individual pairs of radiometers, the cross-power spectra, and 
examine numerous consistency tests. In \S\ref{sec:covariance}, in preparation
for generating  a final combined power spectrum, we present the full covariance
matrix of the cross-power spectra. In \S\ref{sec:combined} we present the
methodology used to produce the final combined power spectrum and its
covariance matrix. In \S\ref{sec:discuss} we compare the \map\ first-year power
spectrum to a compilation of previous CMB measurements and to a prediction
based on a combination of previous CMB data and the 2dFGRS data. We summarize
our results in \S\ref{sec:conclude} and outline the power spectrum data
products being made available through the Legacy Archive for  Microwave
Background Data Analysis (LAMBDA). Appendix~\ref{sec:ps_estimation} reviews two
methods used to estimate the angular power spectrum from CMB maps. 
Appendix~\ref{sec:marginalize}  describes how we account for point source
contamination.   Appendix~\ref{sec:power_opt} presents our approach to
combining multi-channel  data. Appendix~\ref{sec:fisher} describes how the
foreground mask correlates multipole moments in the Fisher matrix, and
Appendix~\ref{sec:wigner} collects some useful properties of the spherical
harmonics.

\section{INSTRUMENTAL PROPERTIES}
\label{sec:instrument}

The \map\ instrument is composed of 10 ``differencing assemblies'' (DAs)
spanning 5 frequencies from 23 to 94 GHz \citep{bennett/etal:2003}. The 2
lowest frequency bands (K and Ka) are primarily Galactic foreground monitors,
while the 3 highest (Q, V, and W) are primarily cosmological bands. There are 8
high frequency differencing assemblies: Q1, Q2, V1, V2, and W1 through W4. 
Each DA is formed from two differential radiometers which are sensitive to 
orthogonal linear polarization modes; the radiometers are designated  1 or 2 
(e.g., V11 or W12) depending on which polarization mode is being sensed.  

The temperature measured on the sky is modified by the properties of the 
instrument. The most important properties that affect the angular power
spectrum are finite resolution and instrument noise. Let $C^{ii'}_l$ denote the
auto or cross-power spectrum evaluated from two sky maps, $i$ and  $i'$, where
$i$ is a DA index.  Further, define the shorthand ${\bf i} \equiv  (i,i')$ to
denote a pair of indices, e.g., (Q1,V2). This spectrum will have the form
\begin{equation}
C^{\bf i}_l = w^{\bf i}_l\,C^{\rm sky}_l + N^{\bf i}_l,
\label{eq:cl_obs}
\end{equation}
where $w^{\bf i}_l \equiv b^i_l\,b^{i'}_l\,p^2_l$ is the window function that
describes the combined smoothing effects of the beam and the finite sky
map pixel size.  Here $b^i_l$ is the beam transfer function for DA $i$, given
by \citet{page/etal:2003b} [note that they reserve the term ``beam window
function'' for $(b^i_l)^2$], and $p_l$ is the pixel transfer function supplied
with the HEALPix package  \citep{gorski/hivon/wandelt:1998}.  $N^{\bf i}_l$ is
the noise spectrum realized in this particular measurement. On average, the
observed spectrum estimates the underlying power spectrum, $C_l$, 
\begin{equation}
\VEV{C^{\bf i}_l} = w^{\bf i}_l\,C_l + \VEV{N^i_l}\, \delta_{ii'}\, ,
\label{eq:cl_obs_bias}
\end{equation}
where $\VEV{N^i_l}$ is the average noise power spectrum for differencing
assembly $i$, and the Kronecker symbol indicates that the noise is uncorrelated
between differencing assemblies. To estimate the underlying power spectrum on
the sky, $C_l$, the  effects of the noise bias and beam convolution must be
removed.  The  determination of transfer functions and noise properties are
thus critical components of any CMB experiment. 

In \S\ref{sec:window_functions} we summarize the results of
\citet{page/etal:2003b} on the \map\ window functions and their uncertainties.
We propagate these uncertainties through to the final Fisher matrix for the 
angular power spectrum. In \S\ref{sec:inst_noise} we present a model of
the \map\ noise  properties appropriate to power spectrum evaluation. For cross
power spectra ($i \ne i'$ above), the noise bias term drops out of 
equation~(\ref{eq:cl_obs_bias}) if the noise  between the two DAs is
uncorrelated.  These cross-power spectra provide a nearly optimal estimate of
the  true power spectrum, essentially independent of errors in the noise model,
thus we use them exclusively in our final power spectrum estimate. The noise
model is primarily used to propagate  noise errors through the analysis, and to
test a variety of different power spectrum estimates for consistency with the
combined cross-power spectrum.

\subsection{Window Functions}
\label{sec:window_functions}

As discussed in \citet{page/etal:2003b}, the instrument beam response was 
mapped in flight using observations of the planet Jupiter.  The signal to noise
ratio is such that the response, relative to the peak of the beam, is measured 
to approximately $-35$ dB in W band, the band with the highest angular
resolution.  The beam widths, measured in flight, range from $0\ddeg82$ at K
band down to $0\ddeg20$ in some of the W band channels (FWHM).  Maps of the
full two-dimensional beam response are  presented in \citet{page/etal:2003b},
and are available with the \map\ first-year data release. The radial beam
profiles obtained  from these maps have been fit to a model consisting of a sum
of Hermite  polynomials that accurately characterize the main Gaussian lobe and
small  deviations from it.  The model profiles are then Legendre transformed to
obtain the beam transfer functions $b^i_l$ for each DA $i$.  Full details of
this  procedure are presented in \citet{page/etal:2003b}, and the resulting
transfer functions are also provided in the first-year data release.  We have 
chosen to normalize the transfer function to 1 at $l=1$ because \map\ 
calibrates its intensity response using the modulation of the CMB dipole ($l =
1$). This effectively partitions calibration uncertainty from window function 
uncertainties.

The beam processing described above provides a straightforward means of 
propagating the noise uncertainty directly from the time-ordered data through
to the final transfer functions.  The result is the covariance matrix 
$\Sigma^i_{b,ll'}$ for the normalized transfer function.  Plots of the diagonal
elements of $\Sigma^i_{b,ll'}$ are presented in \citet{page/etal:2003b}.   The
fractional error in the transfer functions $b^i_l$ are typically 1-2\% in
amplitude.  In the end, these window function uncertainties dominate the small 
off-diagonal elements of the final covariance matrix for the combined
power  spectrum (see \S\ref{sec:covariance}). Additional observations
of Jupiter will  reduce these uncertainties.

An additional source of error in our treatment of the beam response arises from
non-circularity of the main beam.  The effects of  this non-circularity are
mitigated by \map's scan strategy which results in most sky pixels being
observed over a wide range of  azimuth angles.  The effective beam response on
the sky is thus largely  symmetrized.  We estimate that the effects of
imperfect symmetrization produce  window function errors of $<1$\% relative to
a perfectly symmetrized  beam window function
\citep{page/etal:2003b,hinshaw/etal:2003}.  This error is well within the
formal uncertainty given in $\Sigma^i_{b,ll'}$. In \S\ref{sec:combined}  we
infer the optimal power spectrum by combining the 28 cross-power spectra 
measured by the 8 high frequency DAs Q1 through W4.  As part of this process we
marginalize over the window function uncertainty, which automatically 
propagates these errors into the final covariance matrix for the combined power
spectrum.  Both the combined power spectrum and the corresponding Fisher matrix
are part of the first-year data release.

\subsection{Instrument Noise Properties}
\label{sec:inst_noise}

The noise bias term in equation~(\ref{eq:cl_obs_bias}) is the noise per 
$a_{lm}$  mode on the sky. If  auto-power spectra are used in the final power
spectrum estimate, the noise  bias  term must be known very accurately because
it exponentially dominates the  convolved power spectrum at high $l$. If only
cross correlations are used, the noise bias  is only required for propagating
errors. Our final best spectrum is based only on cross correlations, and is
independent of this term.  However, as an independent check of our results, we
evaluate the maximum likelihood spectrum based on a combined Q+V+W sky map. 
The noise bias must be estimated accurately for this application.

In the limit that the time-ordered instrument noise is white, the noise bias is
a constant, independent of $l$. If the time-ordered noise has a $1/f$
component, the bias term will rise at low $l$. In this subsection we  estimate
the noise bias properties for each of the high frequency \map\ radiometers
based on the time-ordered noise properties presented in 
\citet{jarosik/etal:2003b}. While the \map\ radiometer noise is nearly white by
design \citep{jarosik/etal:2003} with $1/f$ knee frequencies of less than 10
mHz for 9 out of 10 differencing assemblies, one of the radiometers (W41) has a
$1/f$  knee frequency of $\sim$45 mHz.   The latter is  large enough that the
deviations of $\VEV{N^i_l}$ from a constant  must be accounted for.

The most reliable way to estimate the effects of $1/f$ noise on the measured 
power spectra is by Monte Carlo simulation.  Using the pipeline simulator
discussed in \citet{hinshaw/etal:2003} we have generated a library of noise 
maps with flight-like properties.  Specifically we have included flight-like
$1/f$ noise in the simulated time-ordered data, and have run each full-year
realization through the map-making  pipeline, including the baseline
pre-whitening discussed in  \citet{hinshaw/etal:2003}.  We evaluate the power
spectra of these maps using the quadratic estimator described in 
Appendix~\ref{sec:ps_estimation} with 3 different pixel weighting schemes. (See
Appendix~\ref{sec:weights} for definitions of the weights, and the $l$ range in
which each is used.) We define the effective noise as a function of $l$ based
on fits to these Monte Carlo noise spectra. For the analyses in this paper,
we fit the spectra to a model of the form
\begin{equation}
\ln \VEV{N^i_l} = \left[\sum_{n=0}^{n_{\rm max}} c^i_n\,(\ln l)^n\right]^{-1}
\label{eq:noise_fit}
\end{equation}
where the $c^i_n$ are fit coefficients given in Table~\ref{tab:noise_fit}, with
$n_{\rm max}=3$ for $l<200$, and $n_{\rm max}=1$ for $l>200$.

\begin{figure}
\begin{center}
\includegraphics[width=0.8\textwidth]{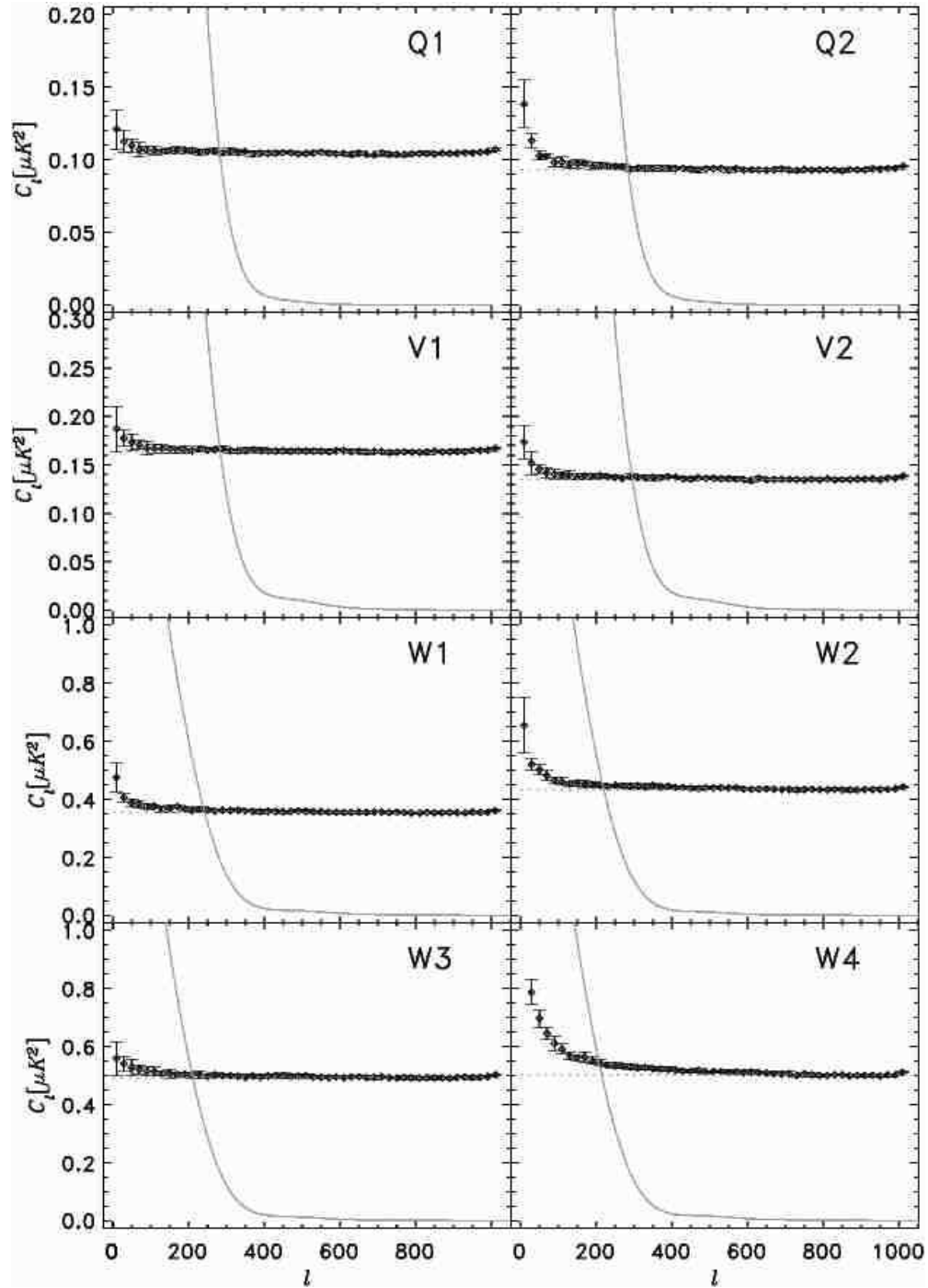}
\end{center}
\caption{The effective noise as a function of $l$ for the 8 differencing
assemblies   used in the combined power spectrum analysis. These spectra were
computed from end-end simulations of noise maps, as discussed in 
\S\ref{sec:inst_noise}.  For illustration, the spectra shown here were computed
using the quadratic estimator with uniform pixel weights.  The actual  noise
model uses three separate weighting schemes in three separate $l$ ranges, and
thus has discontinuities where the effective noise level changes.  The weights
are defined in Appendix~\ref{sec:weights}.
\label{fig:noise_spec}}
\end{figure}

Figure~\ref{fig:noise_spec} shows the noise spectrum derived from the
simulations for each of the 8 high frequency DAs, using uniform weighting over
the entire $l$ range. For comparison, we also plot an estimate of the  CMB
power spectrum from \S\ref{sec:combined} in grey.  Note that the W4 spectrum is
the only one of this set to exhibit deviations from white noise in an $l$
range  where the signal-to-noise is relatively low, and we believe this
simulation slightly over-estimates the $1/f$ noise in the flight W4
differencing assembly \citep{hinshaw/etal:2003b}.

\subsection{Systematic Errors}
\label{sec:systematics}

\citet{hinshaw/etal:2003b} present limits on systematic errors in the
first-year sky maps.  They consider the effects of absolute and relative
calibration errors, artifacts induced by environmental disturbances (thermal
and electrical), errors from the map-making process, pointing errors, and other
miscellaneous effects.  The combined errors due to relative calibration
errors, environmental effects, and map-making errors are limited to $<15$
$\mu$K$^2$ ($2\sigma$) in the quadrupole moment $C_2$ in any of the 8
high-frequency DAs. Tighter limits are placed on higher-order moments. We
conservatively estimate the absolute calibration uncertainty in the first-year
\map\ data to be 0.5\%.

Random pointing errors are accounted for in the beam mapping procedure; the
beam transfer functions  presented by \citet{page/etal:2003b} incorporate
random pointing errors automatically. A systematic pointing error of
$\sim$10\asec at the spin period  is suspected in the quaternion solution that
defines the spacecraft pointing. This is much smaller than the smallest beam
width ($\sim$12\amin at W band), and we estimate that it would produce $<$1\%
error in the angular power spectrum at $l=1000$, thus we do not attempt to
correct for this effect. \citet{barnes/etal:2003} place limits on spurious
contributions due to stray light pickup through the far sidelobes of the
instrument.  They place limits of $<$10 $\mu$K$^2$ on spurious contributions
to $C_l$, at Q through W band, due to far sidelobe pickup.

A detailed model of Galactic foreground emission based on the first-year \map\
data is presented by \citet{bennett/etal:2003c} and is summarized in 
\S\ref{sec:foregrounds}. We show that diffuse foreground emission is a modest
source of contamination at large angular scales ($\gsim$2\dg). {\em Systematic
errors on these angular scales are negligible compared to the (modest) level of
foreground emission.} On smaller angular scale ($\lsim$2\dg), the 1-3\%
uncertainty in the individual beam transfer functions is the largest source of
uncertainty, while for multipole moments greater than $\sim$600, random white
noise from the instrument is the largest source of  uncertainty.

\section{THE DATA}
\label{sec:data}

\begin{figure}
\begin{center}
\includegraphics[width=0.8\textwidth]{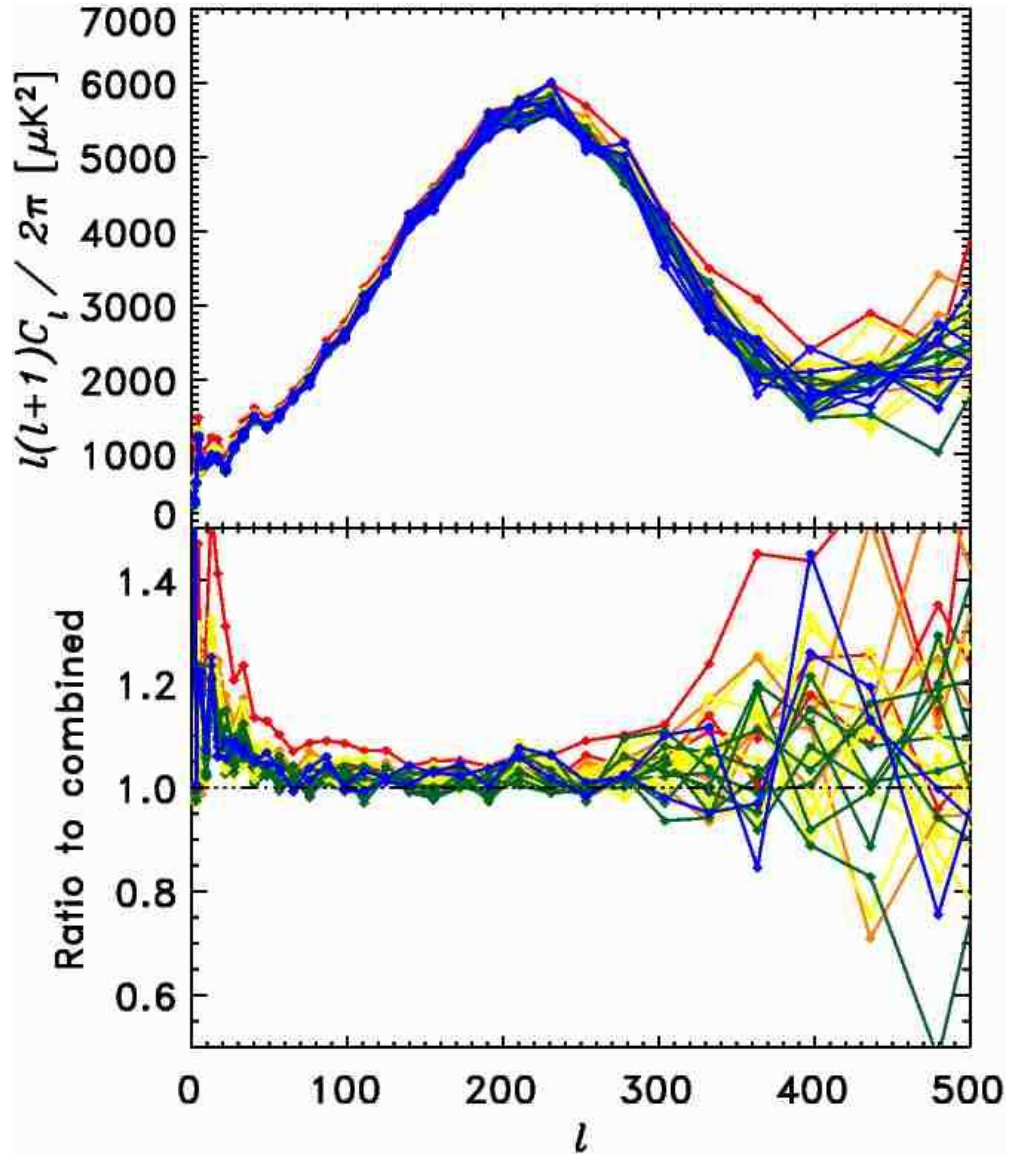}
\end{center}
\caption{The full set of individual cross-power spectra for $l<500$, computed
from the 8 high frequency differencing assemblies Q1 through W4, 28 spectra in
all.  The  spectra were evaluated from the uncorrected sky maps (no Galaxy
model  subtracted) using the Kp2 sky cut with uniform weighting.  The data are
plotted in color by effective frequency $\sqrt{\nu_i\nu_j}$ with red 
corresponding to 41 GHz and blue to 94 GHz. The top panel shows a very robust 
measurement of the first acoustic peak with a maximum near $l \sim 220$.  
There is also a clear indication of the rise to a second peak at  $l \sim 540$
as discussed in \S\ref{sec:discuss}.  The bottom panel shows the ratio of
each channel to the combined spectrum presented in \S\ref{sec:combined}. 
This clearly shows the residual foreground emission due to diffuse Galactic
radio emission at low $l$ and to point sources at higher $l$.  The level of
contamination, which is strongest at Q band, is consistent with the expected
level of foreground emission.  See Figure~\ref{fig:cross_power_corr} for the
spectra after foreground subtraction.
\label{fig:cross_power_raw}}
\end{figure}

Figure~\ref{fig:cross_power_raw} shows the cross-power spectra obtained from all
28 combinations of the 8 differencing assemblies Q1 through W4 using the 
quadratic estimator described in Appendix~\ref{sec:quad_estimation}.  These
spectra have been evaluated with the Kp2 sky cut described in
\citet{bennett/etal:2003c}.  The  spectra are color coded by effective
frequency, $\sqrt{\nu^i\nu^{i'}}$, where  $\nu^i$ is the frequency of
differencing assembly $i$.  The low  frequency (41 GHz) data are shown in red,
the high frequency (94 GHz) data in blue, with intermediate frequencies
following the colors of the rainbow.  The top panel shows $l(l+1) C_l/2\pi$ in
$\mu$K$^2$, while the bottom panel plots the ratio  of each channel to our
final combined spectrum presented in  \S\ref{sec:combined}.  The top panel
shows a very robust measurement of the  first acoustic peak with a maximum near
$l \sim 220$ and a shape that is consistent with the predictions of adiabatic
fluctuation models.  There is also a clear indication of the rise to a second
peak at $l \sim 540$.  See \citet{page/etal:2003c} for an analysis and
discussion of the peaks and troughs in the first-year \map\ power spectrum.

The red data in the top panel show very clearly that the low frequency data 
are contaminated by diffuse Galactic emission at low $l$ and by point sources 
at higher $l$.  The higher frequency data show less apparent contamination, 
consistent with the foreground emission being dominated by radio emission,
rather than thermal dust emission, as expected in this frequency range.

\subsection{Galactic and Extragalactic Foregrounds}
\label{sec:foregrounds}

\citet{bennett/etal:2003c} present a detailed model of the Galactic foreground
emission based on a Maximum Entropy analysis of all 5 \map\ frequency bands, 
in combination with external tracer templates.  They demonstrate that the
emission is well modeled by three distinct emission components. 1) Synchrotron
emission from cosmic  ray electrons, with a steeply falling spectrum in the
\map\ frequency range: $T_A(\nu) \propto \nu^{\beta}$ with $\beta < -3$, 
steepening with increasing frequency.  2) Free-free emission from the 
ionized interstellar medium that is well traced by H$\alpha$ emission in 
regions where the dust extinction is low. 3) Thermal emission from 
interstellar dust grains with an emissivity index $\beta \sim 2.2$.  The model 
has a Galactic signal minimum between V and W band.  

In principal we could subtract the above model from each \map\ channel and 
recompute the power spectrum.  However, since the model is based on \map\ data
that have been smoothed to an angular resolution of $1\ddeg0$, the resulting
maps would have complicated noise properties.  For the purposes of power 
spectrum analysis, we adopt a more straightforward approach based on fitting 
foreground tracer templates to the Q, V, and W band data.  The details of this
procedure, the resulting fit coefficients, and a comparison of the  fits to the
Maximum Entropy model are given in \citet{bennett/etal:2003c}.   They estimate
that the template model removes $\sim$85\% of the foreground  emission in Q, V,
and W bands and that the remaining emission constitutes less than $\sim$2\% of
the CMB variance (up to $l = 200$) in Q band, and $\lsim$1\% of  the CMB
variance in V and W bands.

The contribution from extragalactic radio sources has been analyzed in three 
separate ways.  \citet{bennett/etal:2003c} directly fit for sources in the sky
maps.  The result of this analysis is that we have identified 208 sources in
the \map\ data with sufficient signal to noise ratio to pass the detection
criterion (we estimate that $\sim$5 of these are likely to be spurious).  The 
derived source count law is consistent with the following model for the power 
spectrum of the unresolved sources
\begin{equation}
C_l^{{\bf i},src} = A \left(\frac{\nu_i}{\nu_0}\right)^\beta
\left(\frac{\nu_{i'}}{\nu_0}\right)^\beta w_{l}^{\bf i}, 
\label{eq:source_model}
\end{equation} 
with $A = 0.015$ $\mu$K$^2$ sr (measured in thermodynamic temperature), 
$\beta=-2.0$, and $\nu_0 \equiv 45$ GHz. \citet{komatsu/etal:2003} evaluate the
bispectrum of the \map\ data and are able to fit a non-Gaussian source 
component based on a particular configuration of the bispectrum.  They find 
the same source model, equation~(\ref{eq:source_model}), fits the bispectrum
data.  For the remainder of this section, we adopt this model for correcting
the cross-power spectra.  At Q band (41 GHz) the correction to $l(l+1)C_l/2\pi$
is 868 and 3468 $\mu$K$^2$ at $l$=500 and 1000, respectively.  At W band (94
GHz), the correction is only 31 and 126 $\mu$K$^2$ at the same $l$ values.  For
comparison, the CMB power in this $l$ range is $\sim$ 2000 $\mu$K$^2$. Later,
when we derive a final combined spectrum from the multi-frequency data, we 
adopt equation~(\ref{eq:source_model}) as a model with $A$ as a free parameter.
We simultaneously fit for a combined CMB spectrum and source amplitude and
marginalize over the residual uncertainty in A.  The best-fit source amplitude
from this process is consistent with the other two methods.

\begin{figure}
\begin{center}
\includegraphics[width=0.8\textwidth]{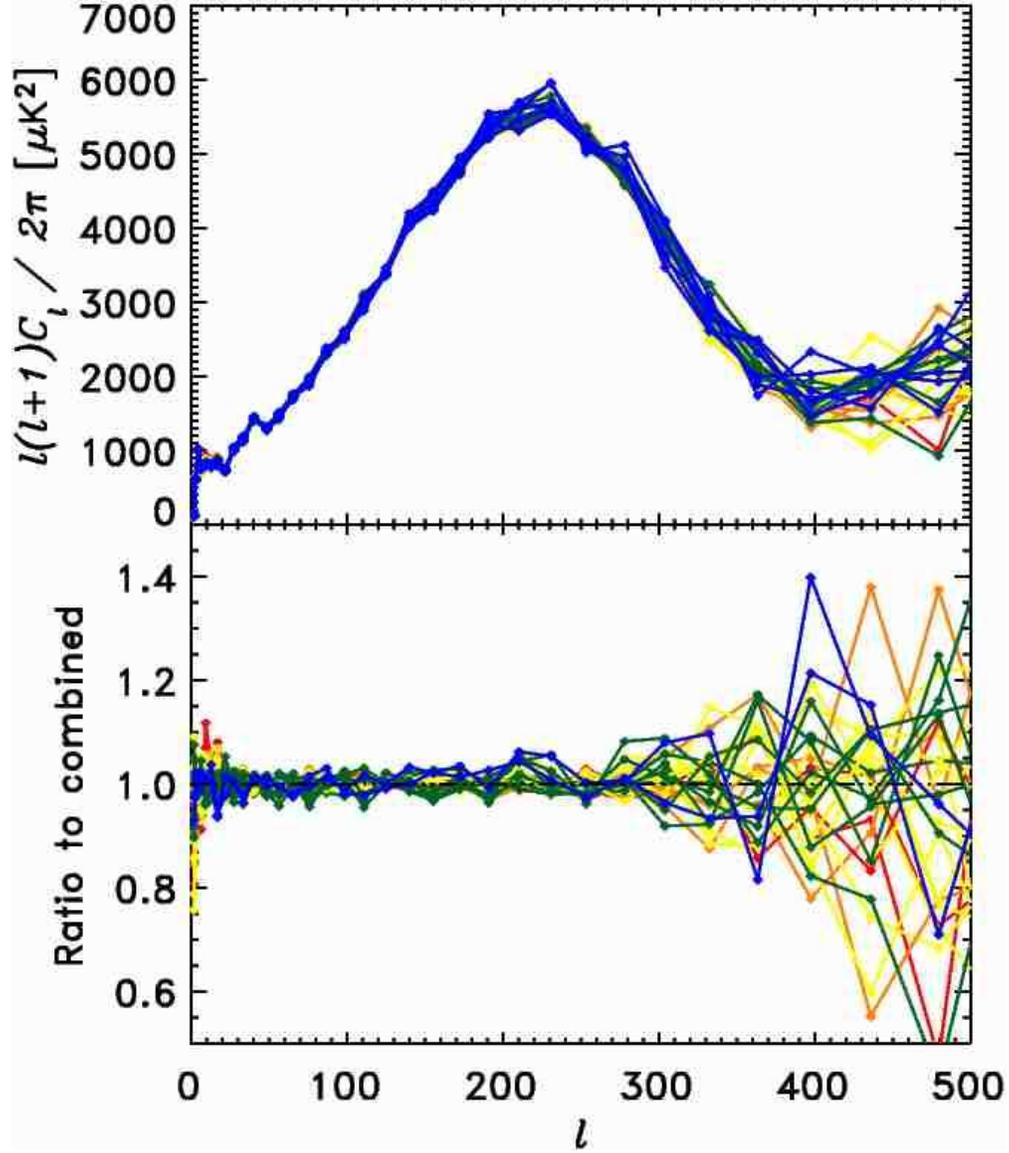}
\end{center}
\caption{The same set of cross-power spectra as shown in
Figure~\ref{fig:cross_power_raw}.  Here, the foreground model discussed in 
\S\ref{sec:foregrounds} has been subtracted from each channel. The bottom panel
shows the ratio of each of the 28 cross-power spectra to the combined spectrum
presented in \S\ref{sec:combined}.  Aside from a $\sim$10\% discrepancy in the
Q band data at $l < 20$, the data are consistent with  each other to the
sensitivity limits of the individual spectra.  Because the  \map\ data are not
sensitivity limited at low $l$, we use only V and W band  data in the final
combined spectrum for $l < 100$ (see \S\ref{sec:combined}) to minimize residual 
Galactic contamination.
\label{fig:cross_power_corr}}
\end{figure}

Figure~\ref{fig:cross_power_corr} shows the cross-power spectra obtained 
from the same 28 combinations as in Figure~\ref{fig:cross_power_raw}, this 
time with the Galactic template model and source model subtracted.  The bottom
panel of the Figure shows the ratio of the 28 channels to the combined spectrum
obtained in \S\ref{sec:combined}.  The 28 cross-power spectra are  consistent
with each other at the 5 to 20\% level over the $l$ range  2 -- 500. Similar
scatter is seen in Monte Carlo simulations of an ensemble of 28 cross-power
spectra with \map's beam and noise properties.  The only significant deviation
lies in the Q band data at low $l$ which is $\sim$10\%  below the higher
frequency bands at $l < 20$. This is consistent with the accuracy estimated
above for the Galactic template model, see also Figure 11 of 
\citet{bennett/etal:2003c} for images of the maps after Galactic template
subtraction.  Since the \map\ data are not noise limited at low $l$, we use
only V and W band data in the final combined spectrum for $l < 100$.

The subtraction of the source model, equation~(\ref{eq:source_model}), brings the  Q
band spectrum into good agreement with the other cross-power spectra up to  $l
\sim 500$.  At higher $l$, the Q band data contributes very little to  the
final combined spectrum because the (normalized) Q band window function  has
dropped to less than 5\% \citep{page/etal:2003b}.  As discussed in \S
\ref{sec:combined}, we marginalize  over the source amplitude uncertainty,
$\delta A$, when obtaining the final power spectrum estimate and associated
covariance  matrix.  Thus the uncertainty is also accounted for in 
subsequent cosmological parameter fits \citep{verde/etal:2003,spergel/etal:2003,
peiris/etal:2003}.

\begin{figure}
\begin{center}
\includegraphics[width=0.8\textwidth]{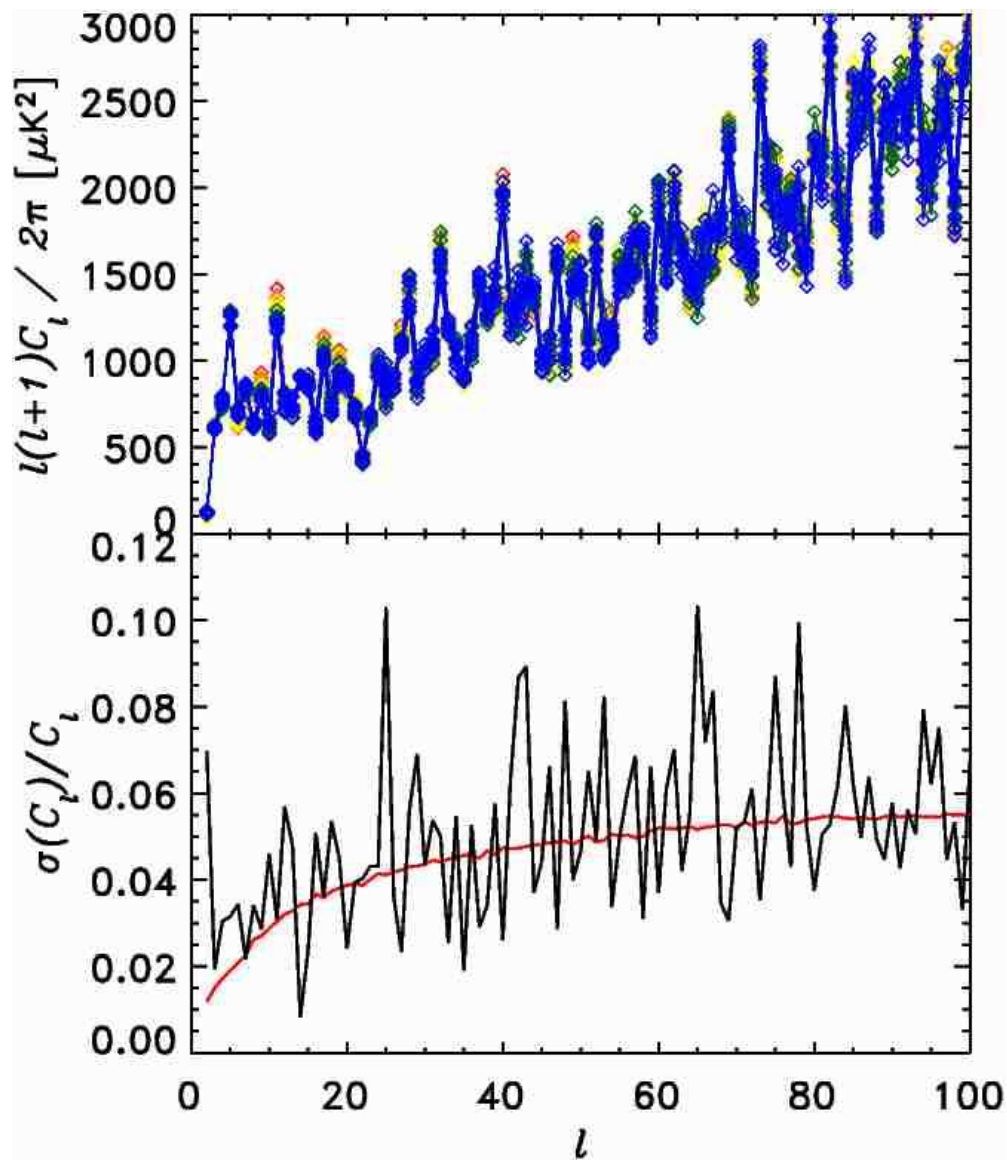}
\end{center}
\caption{The same set of cross-power spectra as shown in
Figure~\ref{fig:cross_power_corr}, but showing the low $l$ spectrum unbinned. The
agreement between the individual spectra is striking. The low value of the
quadrupole moment, $C_2$, that was first seen by {\sl COBE-DMR} is also seen in
the \map\ data. The steep, nearly linear rise in the spectrum from $l=2$ to  5
translates to a near absence of power in the angular correlation function at
separations larger than $\sim$60$^{\circ}$ \citep{spergel/etal:2003,
bennett/etal:2003b}. This was also seen in the {\sl COBE-DMR} data, but it is
now clear that this is not due to Galaxy modeling errors. The bottom panel
shows the fractional rms among the 28 \map\ cross-power spectra in black,
while the red curve shows the same statistic averaged over an ensemble of 1000
Monte Carlo realizations.  Based on this we estimate the measurement error on
the combined spectrum to be $<$2-5\% for $l < 100$.
\label{fig:cross_power_corr_100}}
\end{figure}

Figure~\ref{fig:cross_power_corr_100} shows a close-up of the 28 cross-power 
spectra in Figure~\ref{fig:cross_power_corr} up to $l=100$.  The top panel
shows the raw (unbinned) data which has correlations of $<$2\% between 
neighboring points is this $l$ range (see \S\ref{sec:covariance}). These
data are strikingly consistent with each other and support the conclusion that 
systematic errors at low $l$ are insignificant.  To assess the level of scatter
that does exist between the spectra, we have generated a Monte Carlo simulation
in which we compute the rms scatter among the 28 spectra at each $l$,
relative to the measured power.  The bottom panel of
Figure~\ref{fig:cross_power_corr_100} shows the results of this simulation,
averaged over 1000 realizations, compared to the relative rms scatter in the
data.  The agreement is excellent, indicating that the uncertainty in the
measured power spectrum in this $l$ range is a few percent and is consistent
with a combination of instrument noise and mode coupling due to the 15\% sky
cut.

Another striking feature is the low amplitude of the observed quadrupole, and
the sharp rise in power, almost linear in $l$, to $l=5$. 
\citet{bennett/etal:2003b} quote a value for the rms quadrupole amplitude,
$Q_{rms} = \sqrt{(5/4\pi)C_2} = 8 \pm 2$ $\mu$K, where the uncertainty is
largely due to Galactic model uncertainty.  This is consistent with the
amplitude measured by the {\sl COBE-DMR} experiment, $Q_{rms} = 10^{+7}_{-4}$
$\mu$K \citep{bennett/etal:1996}.  The fast, nearly linear rise to $l=5$
produces an angular correlation function with essentially no power on angular
scales  $\gsim$60\dg, again in excellent agreement with the {\sl COBE-DMR}
correlation function \citep{bennett/etal:2003b,hinshaw/etal:1996b}.  In the
context of a standard $\Lambda$CDM model, the probability of observing this
little power on scales greater than 60\dg\ is $\sim 2 \times 10^{-3}$ 
\citep{bennett/etal:2003b}.

\section{THE FULL COVARIANCE MATRIX}
\label{sec:covariance}

In \S\ref{sec:combined} we derive our best estimate of the angular power
spectrum by optimally combining the 28 cross-power spectra discussed above. 
The procedure for combining spectra requires the full covariance matrix of the
individual cross-power spectra -- in this section we outline the salient
features of this matrix.  There are six principal sources of variance for the
measured spectra, $C^{\bf i}_l$: cosmic variance,  instrument noise, mode
coupling due to the foreground mask, point source subtraction errors,
uncertainty in the beam window functions, and an overall calibration
uncertainty. We ignore uncertainties in the diffuse foreground correction since
they are everywhere sub-dominant to the cosmic variance uncertainty (see
\S\ref{sec:foregrounds}).

We may write the covariance matrix as
\begin{equation}
(\Sigma_{\rm full})^{\bf ij}_{ll'} = \left<
  [ C^{\bf i}_l - (C_l + AS^{\bf i})w^{\bf i}_l\,]
  [ C^{\bf j}_{l'} - (C_{l'} + AS^{\bf j})w^{\bf j}_{l'}\,]
  \right>,
\label{eq:cov_ci_l}
\end{equation}
where the angle brackets represent an ensemble average, $C_l$ is the true 
underlying power spectrum, $w^{\bf i}_l$ is the window function of spectrum
${\bf i}$, and $AS^{\bf i}$ is the point source contribution  to spectrum ${\bf
i}$.  Here we have defined a point source spectral function, $S^{\bf i}$, as
\begin{equation}
S^{\bf i} \equiv \left(\frac{\nu_i}{\nu_0}\right)^\beta
                 \left(\frac{\nu_{i'}}{\nu_0}\right)^\beta,
\label{eq:spec_model}
\end{equation}
where $\nu_i$, $\nu_0$, and $\beta$ are as defined after 
equation~(\ref{eq:source_model}).  Note that $(\Sigma_{\rm full})^{\bf
ij}_{ll'}$ is  symmetric in both $(ll')$ and $({\bf ij})$.

In the process of forming the combined spectrum we will estimate a best-fit
point source amplitude, $\bar{A}$, and subtract the corresponding source
contribution from each spectrum ${\bf i}$. We thus rewrite $\Sigma_{\rm full}$ as
\begin{equation}
(\Sigma_{\rm full})^{\bf ij}_{ll'} = \left<
  [ \bar{C}^{\bf i}_l - (C_l + \delta AS^{\bf i})w^{\bf i}_l\,]
  [ \bar{C}^{\bf j}_{l'} - (C_{l'} + \delta AS^{\bf j})w^{\bf j}_{l'}\,]
  \right>,
\label{eq:cov_ci_src}
\end{equation}
where $\bar{C}^{\bf i}_l \equiv C^{\bf i}_l - \bar{A}S^{\bf i}w^{\bf i}_l$ is 
the source-subtracted spectrum, and $\delta A \equiv A-\bar{A}$ is the residual
source amplitude, which we will marginalize over as a nuisance parameter.

We may expand the covariance matrix as
\begin{equation}
\Sigma_{\rm full} = \Sigma_{\rm cv} + \Sigma_{\rm mask} + \Sigma_{\rm src}
  + \Sigma_{\rm win}.
\label{eq:cov_3}
\end{equation}
We discuss each of these contributions in more detail below.

{\em Cosmic Variance, Noise, and Mode Coupling}.--- 
The first two terms, $\Sigma_{\rm cv}+\Sigma_{\rm mask}$, incorporate the
combined uncertainty due to cosmic variance, instrument noise, and mode
coupling due to the foreground mask,
\begin{equation}
(\Sigma_{\rm cv}+\Sigma_{\rm mask})^{\bf ij}_{ll'} = \left<
  [ \bar{C}^{\bf i}_l - C_l\,w^{\bf i}_l\,]
  [ \bar{C}^{\bf j}_{l'} - C_{l'}\,w^{\bf j}_{l'}\,]
  \right>,
\label{eq:cov_ci_cut}
\end{equation}
where $w^{\bf i}_l$ is fixed at its measured value.  We have split this
contribution into two pieces to mimic the procedure we actually use to compute
the covariance matrix. As outlined in  \S\ref{sec:opt_combined}, we start with
$\Sigma_{\rm cv}$, then incorporate the effects of point source error and
window function error.  We do not add  the effects of mode coupling,
$\Sigma_{\rm mask}$, until the very end of the computation. We consider this
term in  more detail in Appendix~\ref{sec:fisher}. 

{\em Point Source Subtraction Errors}.--- 
The third term, $\Sigma_{\rm src}$, is due to uncertainty in the point source 
amplitude determination
\begin{equation}
(\Sigma_{\rm src})^{\bf ij}_{ll'} = 
  S^{\bf i} w^{\bf i}_l \, \sigma_{\rm src}^2 \, w^{\bf j}_{l'} S^{\bf j},
\label{eq:AS}
\end{equation}
where $\sigma_{\rm src}^2\equiv \left<(A-\bar{A})^2\right>$ is the variance  in
the best-fit amplitude $\bar{A}$, and we assume that the frequency dependence, 
$S^{\bf i}$, is perfectly known.  In practice, we do not explicitly evaluate
$\Sigma_{\rm src}$ as given above, rather we employ a method based on 
marginalizing a Gaussian likelihood function, ${\cal L}(C^{\bf i}_l,A)$, over a
nuisance parameter $A$.  This process, which is discussed in
Appendix~\ref{sec:marginalize}, effectively yields  $\Sigma_{\rm cv} +
\Sigma_{\rm src}$.

{\em Window Function and Calibration Errors}.--- 
The fourth term, $\Sigma_{\rm win}$, is due to uncertainty in the beam window 
function, $w^{\bf i}_l$.  This term arises from fluctuations in the window
function which cause the measured spectrum, $C^{\bf i}_l$ to differ from our
estimate of the convolved spectrum, $C_lw^{\bf i}_l$, where $w^{\bf i}_l$ is 
the estimated window function.  This contribution has the form
\begin{equation}
(\Sigma_{\rm win})^{\bf ij}_{ll'}\,\, = \,\,
  C_l\,\left< \Delta w^{\bf i}_l\cdot\Delta w^{\bf j}_{l'} \right>\,C_{l'}.
  \label{eq:b}
\end{equation}
Recall from \S\ref{sec:instrument} that $w^{\bf i}_l = b^i_lb^{i'}_lp^2_l$
where $b^i_l$ is the beam transfer function for DA $i$ and $p^2_l$ is the pixel
window function.  Define $u^i_l \equiv \Delta b^i_l/b^i_l$ to be the fractional 
error in $b^i_l$, then to first order in $u^i_l$ we have
\begin{equation}
\left< \Delta w^{\bf i}_l\cdot\Delta w^{\bf j}_{l'} \right> = 
w^{\bf i}_l\,\left< u^{i}_l u^{j}_{l'} + u^{i}_l u^{j'}_{l'}
                  + u^{i'}_l u^{j}_{l'} + u^{i'}_l u^{j'}_{l'}
 \right>\,w^{\bf j}_{l'}\,.
\end{equation}
For \map\ the uncertainty in $b^i_l$ is uncorrelated between DAs, thus the
above expression reduces to
\begin{equation}
\left< \Delta w^{\bf i}_l\cdot\Delta w^{\bf j}_{l'} \right> = 
w^{\bf i}_l\,\left[ \Sigma^i_{{\rm u},{ll'}}(\delta_{ij} + \delta_{ij'})
  + \Sigma^{i'}_{{\rm u},{ll'}}(\delta_{i'j} + \delta_{i'j'}) \right]\,w^{\bf j}_{l'}\,,
\end{equation}
where we have defined
\begin{equation}
\Sigma^i_{{\rm u},{ll'}} \equiv \left< u^i_l u^i_{l'} \right> 
 = (b^i_lb^i_{l'})^{-1}\,\Sigma^i_{{\rm b},{ll'}}\,,
\end{equation}
and where $\Sigma^i_{{\rm b},{ll'}}$ is the covariance matrix of the beam
transfer function for DA $i$ given by \citet{page/etal:2003b}.  When generating
the combined spectrum in the next section, we add $\Sigma_{\rm win}$ to the
above contributions, giving $\Sigma_{\rm cv} + \Sigma_{\rm src} + 
\Sigma_{\rm win}$.

The \map\ absolute calibration uncertainty is 0.5\%.  We do not explicitly
incorporate this contribution in the covariance matrix. Instead, we propagate a
0.5\% uncertainty into the normalization of the final power spectrum amplitude 
\citep{spergel/etal:2003}.

\section{THE COMBINED POWER SPECTRUM}
\label{sec:combined}

In \S\ref{sec:data} we demonstrated that the three high frequency bands of
\map\ data produced consistent estimates of the angular power spectrum, after a
modest correction for diffuse Galactic emission and extragalactic point
sources.  It is therefore justifiable to combine these data into a single
``optimal'' estimate of the angular power spectrum of the CMB.  In this
section, we provide an overview of two methods we use to generate a single
combined spectrum.  The first is a multi-step process that simultaneously fits
the 28  cross-power spectra presented above to a single CMB power spectrum and
a point source model, equation~(\ref{eq:source_model}), while correctly
propagating beam and residual point source uncertainties through to a final
Fisher matrix.  This spectrum constitutes our best estimate of the CMB power
spectrum from the first-year \map\ data. The second spectrum, which serves as a
cross check of the first, is based on forming a single co-added sky map from
the Q1 through W4 maps, and using the quadratic estimator with noise  bias
subtraction.  We compare the two spectra in \S\ref{sec:compare}.

\subsection{Method I -- Optimal Combination of Cross-Power Spectra}
\label{sec:opt_combined}

Since this method is relatively complicated, we outline the basic procedure
here and relegate the details to Appendices, as indicated. We present the
result in \S\ref{sec:compare}.  The steps are as follows.
\begin{enumerate}

\item Subtract best-fit Galactic foreground templates from each of the maps
Q1 through W4, using the coefficients given in Table~3 of \citet{bennett/etal:2003c}.

\item Evaluate the 28 cross-power spectra from the maps Q1 through W4, where
each spectrum has been evaluated using the quadratic estimator of 
Appendix~\ref{sec:quad_estimation} with the weighting scheme defined in 
Appendix~\ref{sec:weights}.  

\item Collect the noise bias estimate, $\left< N^i_l\right>$, for each DA from 
\S\ref{sec:inst_noise}. These estimates are used in the calculation of the
covariance  matrix for the combined spectrum, and in setting the relative weight of
each cross-power spectrum in the final combined spectrum.

\item Apply the procedure presented in Appendix~\ref{sec:amp_ps} to obtain an
estimate for the point source amplitude.  The value obtained is 
$\bar{A}=0.0155 \pm 0.0017$, roughly independent of $l_{\rm max}$ in 
equation~(\ref{eq:ps_amplitude}).  This value is in good agreement with an
estimate based on the bispectrum \citep{komatsu/etal:2003}, and on an
extrapolation of point source counts \citep{bennett/etal:2003b}. Subtract the
point source contribution from the cross-power spectra: $\bar{C}^{\bf i}_l
= C^{\bf i}_l - \bar{A}S^{\bf i}w^{\bf i}_l$.

\item Compute an approximate form of the full covariance matrix discussed in
\S\ref{sec:covariance}, $\tilde{\Sigma}_{\rm full}$. The procedure we use
produces a covariance matrix which includes cosmic variance, instrument noise,
source subtraction uncertainties, and window function uncertainties.  At this
stage of the  process, it does  not yet include the effects of mode coupling. 
More  details are given in \S\ref{sec:covariance} and 
Appendix~\ref{sec:marginalize}.

\item Invert the approximate covariance matrix for use in computing the optimal
spectrum.  This is the most computationally intensive step in the process.

\item Compute the final combined spectrum from the 28 $C^{\bf i}_l$ as per the
procedure given in Appendix~\ref{sec:power_opt}. In particular, assume a
fiducial cosmological model (as specified in the Appendix), and use
equation~(\ref{eq:cl_comb}), with  $\Sigma_{{\rm full} \; l l'}^{\bf ij} =
\tilde\Sigma_{{\rm full} \;l l'}^{\bf ij}$.  This produces a final spectrum
which is very nearly optimal.

Note: for $l<100$ we use a surrogate  procedure for computing the combined
spectrum.  In order to minimize Galactic foreground contamination, we use only
V and W band data.  Moreover, because the statistics of the  $C_l$ are mildly
non-Gaussian, and because point source contamination and window function
uncertainties are small, the ``optimal'' machinery developed above is
unnecessary. Rather we simply form a weighted average spectrum from the V and W
band $C^{\bf i}_l$.

\item Compute the approximate inverse-covariance matrix, $\tilde{Q}$, for the
combined  spectrum using equation~(\ref{eq:qll_comb}). This matrix is approximate
in  two ways. 1) It does not yet incorporate the effects of mode coupling --
this  is added below. 2) It has been evaluated for a fiducial cosmological
model, $\tilde{Q}(C^{\rm fid}_l)$, while in a likelihood application, we need
to evaluate  $\tilde{Q}(C^{\rm th}_l)$ for an arbitrary model, $C^{\rm th}_l$
-- we add this next. 

\item Introduce the dependence on cosmological model into $\tilde{Q}$ as
follows. Invert $\tilde{Q}$ to obtain the approximate covariance matrix of the
combined spectrum, $\tilde{\Sigma}$. The off-diagonal terms of $\tilde{\Sigma}$
are small and weakly dependent on cosmological model, so we expand
$\tilde{\Sigma}$ as
\begin{equation}
\tilde{\Sigma}_{ll'} = D_l \delta_{ll'} + \epsilon_{ll'},
\label{eq:def_eps}
\end{equation}
where $\epsilon_{ll'}$ encodes the mode coupling due to window function and
source subtraction uncertainties. This relation defines $\epsilon_{ll'}$, which
we take to be zero on the diagonal.  $D_l$ is dominated by cosmic variance and
noise;  in order to separate the two contributions, we introduce an {\it ansatz} 
of the form 
\begin{equation}
D_l \equiv \frac{2}{2l+1}\frac{1}{f_{\rm sky}}(C^{\rm fid}_l + N^{\rm eff}_l)^2,
\label{eq:ansatz_dl}
\end{equation} 
where $C^{\rm fid}_l$ is the fiducial model used to generate the combined 
spectrum (see Appendix~\ref{sec:power_opt}), and $N^{\rm eff}_l$ is the
effective noise in the combined spectrum, which is defined by this equation.
The dependence on cosmological model is introduced in the covariance
matrix by re-computing $D_l$ with $C^{\rm fid}_l \to C^{\rm th}_l$, leaving
$N^{\rm eff}_l$ fixed.

\item Estimate the coupling induced by the foreground mask as described in
Appendix~\ref{sec:fisher}.  The effect of the mask on the off-diagonal 
elements of the Fisher (inverse-covariance) matrix can be written as
\begin{equation}
(\Sigma^{-1}_{\rm mask})_{ll'} \equiv  r_{ll'}\,(D'_l D'_{l'})^{-1/2},
\end{equation}
where $D'_l \equiv D_l f^{-1}_{\rm sky}$. This expression parameterizes, and
defines, the off-diagonal mode coupling in the form of a correlation matrix,
$r_{ll'}$. Note that $\tilde F_{ll'}$, defined in Appendix
\ref{sec:fisher_interpol}, is related to $r_{ll'}$ by $\tilde F_{ll'}\equiv
\delta_{ll'} + r_{ll'}$.

\item The final Fisher (curvature) matrix, $Q_{ll'}$, is obtained using 
\begin{equation}
Q_{ll'} = (D'_l)^{-1} \delta_{ll'} - \epsilon_{ll'}\,(D'_lD'_{l'})^{1/2}
 + r_{ll'}\,(D'_l D'_{l'})^{-1/2},
\end{equation}

\end{enumerate}

We further calibrate $D'_{l}$ with Monte Carlo simulations. This calibration
process, and a description of how the curvature matrix is used in a maximum
likelihood  determination of cosmological parameters, is given in \S2 of
\citet{verde/etal:2003}. As part of the first-year data release, we provide a
Fortran 90 subroutine to evaluate the likelihood of a given cosmological model,
$C^{\rm th}_l$, given the \map\ data (supplied in the routine).  The code also
optionally returns the Fisher (inverse-covariance) matrix for the combined
spectrum. 

\subsection{Method II -- Combined Sky Map}
\label{sec:map_combined}

\begin{figure}
\begin{center}
\includegraphics[angle=90,width=1.0\textwidth]{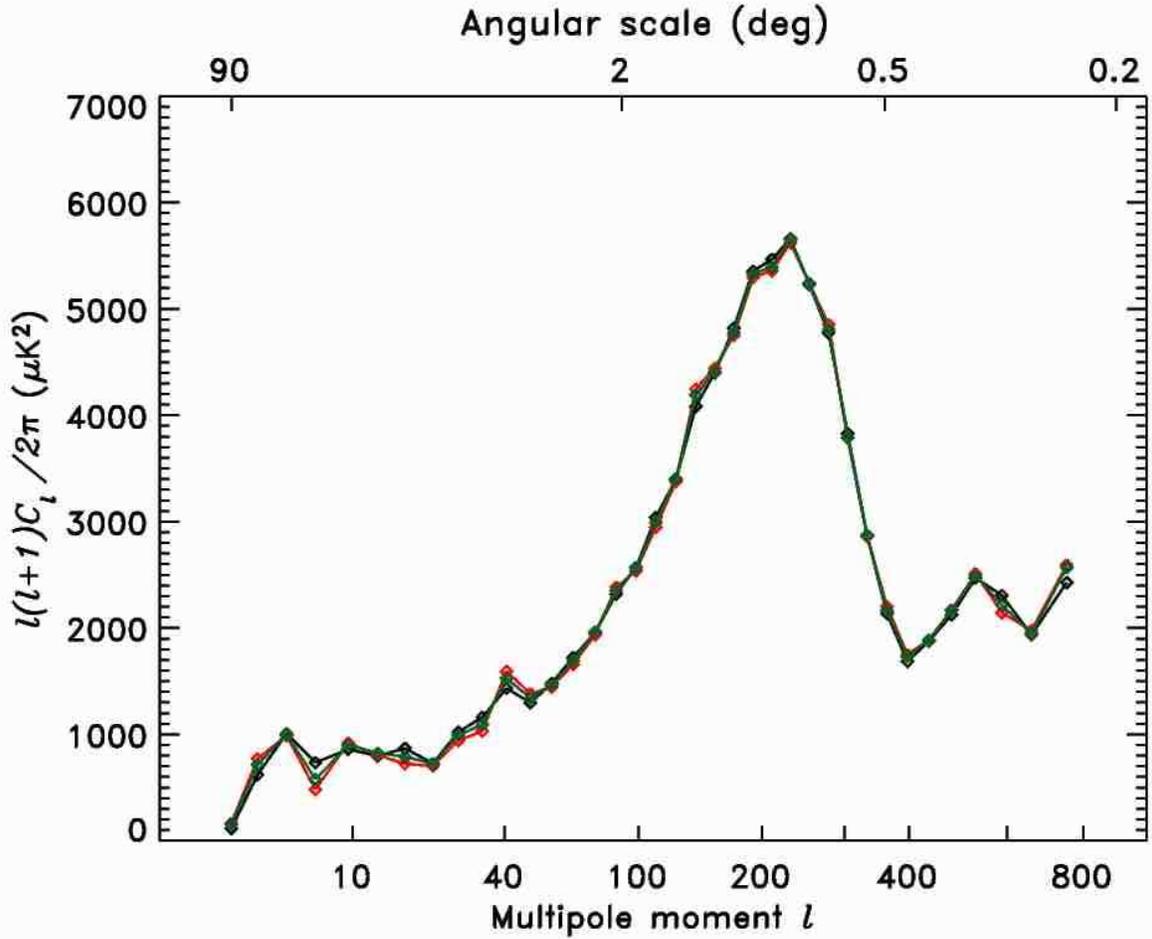}
\end{center}
\caption{The auto-power spectrum of the combined Q+V+W map evaluated with the
three weighting schemes defined in Appendix~\ref{sec:weights}. In each case,
the spectrum was computed over the entire $l$ range, and black shows
uniform weights, red shows inverse-noise weights, and green shows transitional
weights. The agreement is excellent.
\label{fig:cf_weights}}
\end{figure}

Our second approach is to form a single co-added map from the Q1 through W4
maps,
\begin{equation}
T = \frac{\sum_{i=3}^{10} T'_i/\sigma^2_{0,i}}{\sum_{i=3}^{10} 1/\sigma^2_{0,i}},
\label{eq:comb_map}
\end{equation}
where $T'_i$ is the sky map for DA $i$ with the best-fit Galactic template
model subtracted, and $\sigma^2_{0,i}$ is the noise per observation for DA $i$,
given by  \citet{bennett/etal:2003b}.  We evaluate the power spectrum of this
map on the Kp2 cut sky using the quadratic estimator in 
Appendix~\ref{sec:quad_estimation}. An effective noise model is obtained by
using the same approach as described in \S\ref{sec:inst_noise}: we generate
co-added noise maps from the library of end-to-end simulations, evaluate their
average spectra, then fit a noise model.  The noise bias model is then
subtracted from the power spectrum of the combined temperature  map.  We have
performed this analysis with 3 distinct pixel weighting schemes (see
Appendix~\ref{sec:weights}) and 3 corresponding noise models.  The results are
shown in Figure~\ref{fig:cf_weights} where it is seen that the three cases are
virtually indistinguishable.

In effect, this analysis uses both the auto- and cross-power spectra.  We view
this as a useful check of the more sophisticated procedure described in 
\S\ref{sec:opt_combined}, but we do not rely on it for a final result.
Uncertainties in the noise model only effect the fourth moment of the
cross-power spectra, but they effect the second moment of the auto-power
spectra and potentially bias the final result. The $\sim$6\% sensitivity
advantage gained by including auto-power spectra was not deemed worth the 
effort required to guarantee that the final result was not biased.

\subsection{Comparison of Results}
\label{sec:compare}

\begin{figure}
\begin{center}
\includegraphics[angle=90,width=1.0\textwidth]{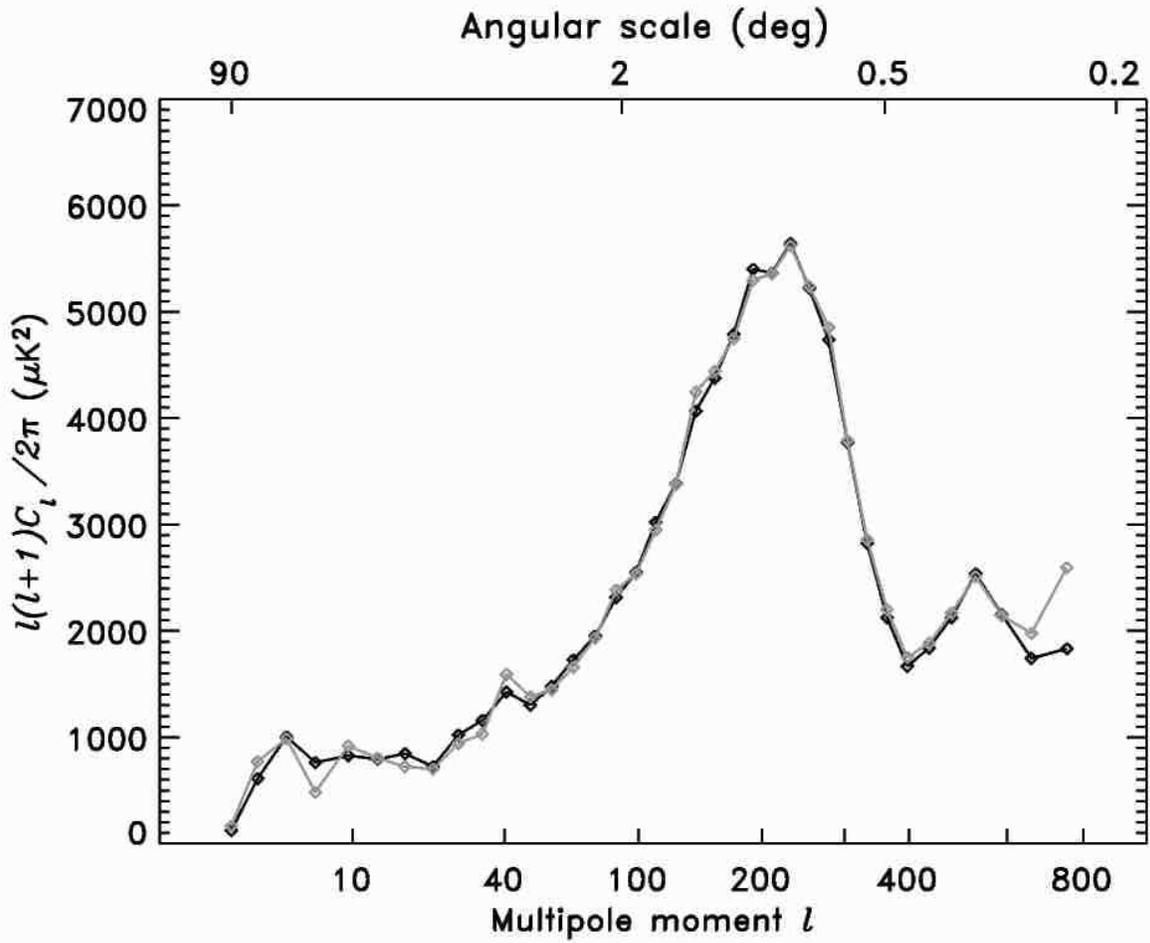}
\end{center}
\caption{This figure compares the auto-power spectrum computed from the
combined Q+V+W map (grey) to the optimally combined cross-power spectrum
(black).  The close agreement indicates that the noise properties of the
first-year \map\ data are well understood
\label{fig:cf_combined}}
\end{figure}

Figure~\ref{fig:cf_combined} compares the power spectra obtained from methods I
and II above.  The combined cross-power spectrum from \S\ref{sec:opt_combined}
is shown in black, the  auto-power spectrum obtained in
\S\ref{sec:map_combined} from the co-added map is shown in grey.  The two 
methods agree extremely well with the only notable deviation being at the
highest $l$ range probed by the first-year data.  This is the regime where the
auto-power spectrum will be most sensitive to the noise bias subtraction. As
can be seen in the error estimates shown in Figure~\ref{fig:final_spec}, the
deviation is less than 1$\sigma$.

\begin{figure}
\begin{center}
\includegraphics[angle=90,width=1.0\textwidth]{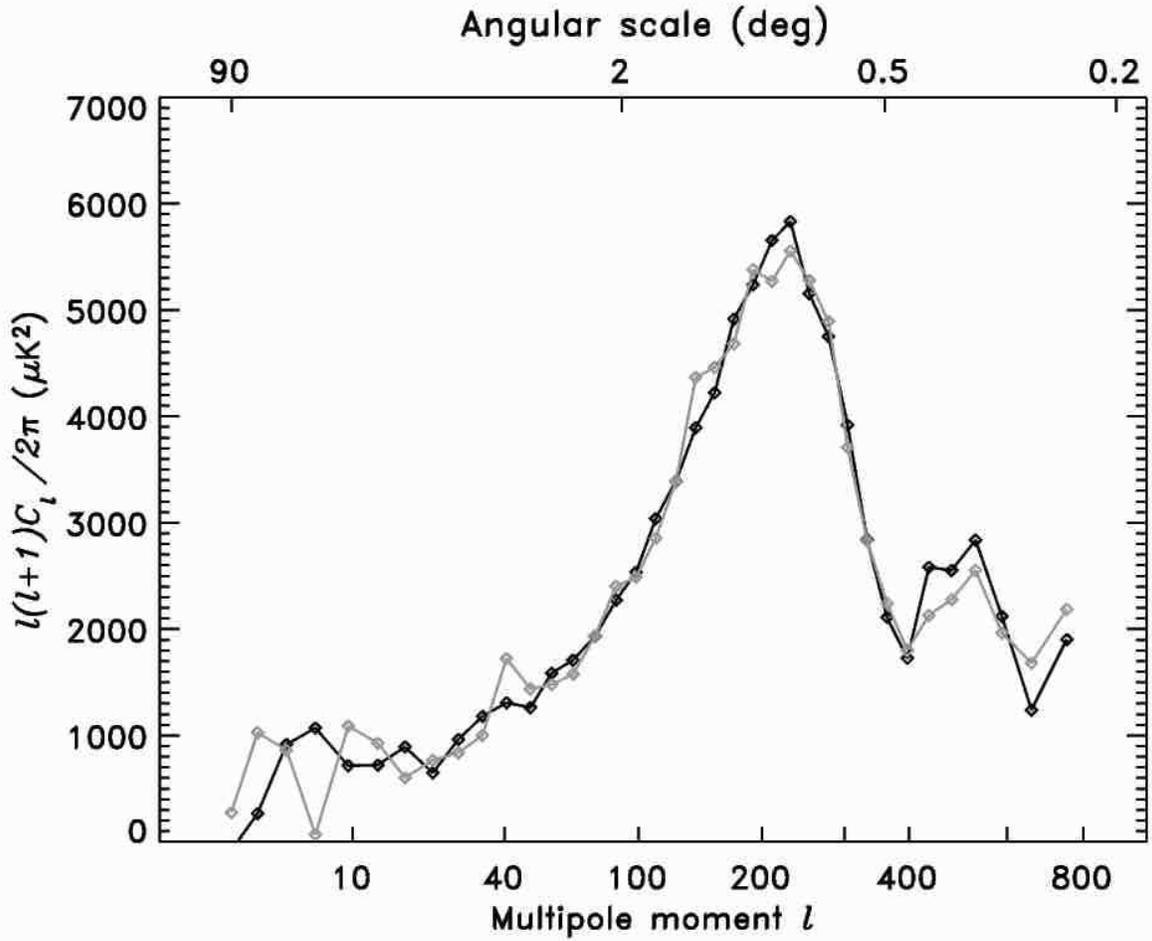}
\end{center}
\caption{A comparison of the power spectrum computed with data from the 
ecliptic plane (black) vs. data from the ecliptic poles (grey).  Note that
some of the ``bite'' features that appear in the combined spectrum are not
robust to data excision.  There is also no evidence that beam ellipticity, 
which would be more manifest in the plane than in the poles, systematically
biases the spectrum.  This is consistent with estimates of the effect given
by \citet{page/etal:2003b}.
\label{fig:cf_plane_pole}}
\end{figure}

A separate test of robustness is to compute the angular power spectrum in
separate regions of the sky to see if the spectrum changes. We have computed
the power spectrum in two subsets of the sky -- the ecliptic poles, and the
ecliptic plane, using the quadratic estimator with the combined sky map.  The
results are shown in  Figure~\ref{fig:cf_plane_pole} where the pole data is
shown in grey and the  plane data in black. The two spectra are very consistent
overall, but some of the features that appear in the combined spectrum,  such
as the ``peak'' at $l\sim 40$ and the ``bite'' at $l \sim 210$, are not robust
to this test, thus we consider these features to be of marginal significance.
There is also no evidence that beam ellipticity,  which would be more manifest
in the plane than in the poles, systematically biases the spectrum.  This is
consistent with estimates of the effect given by \citet{page/etal:2003b}.

\section{DISCUSSION}
\label{sec:discuss}

\begin{figure}
\begin{center}
\includegraphics[angle=90,width=1.0\textwidth]{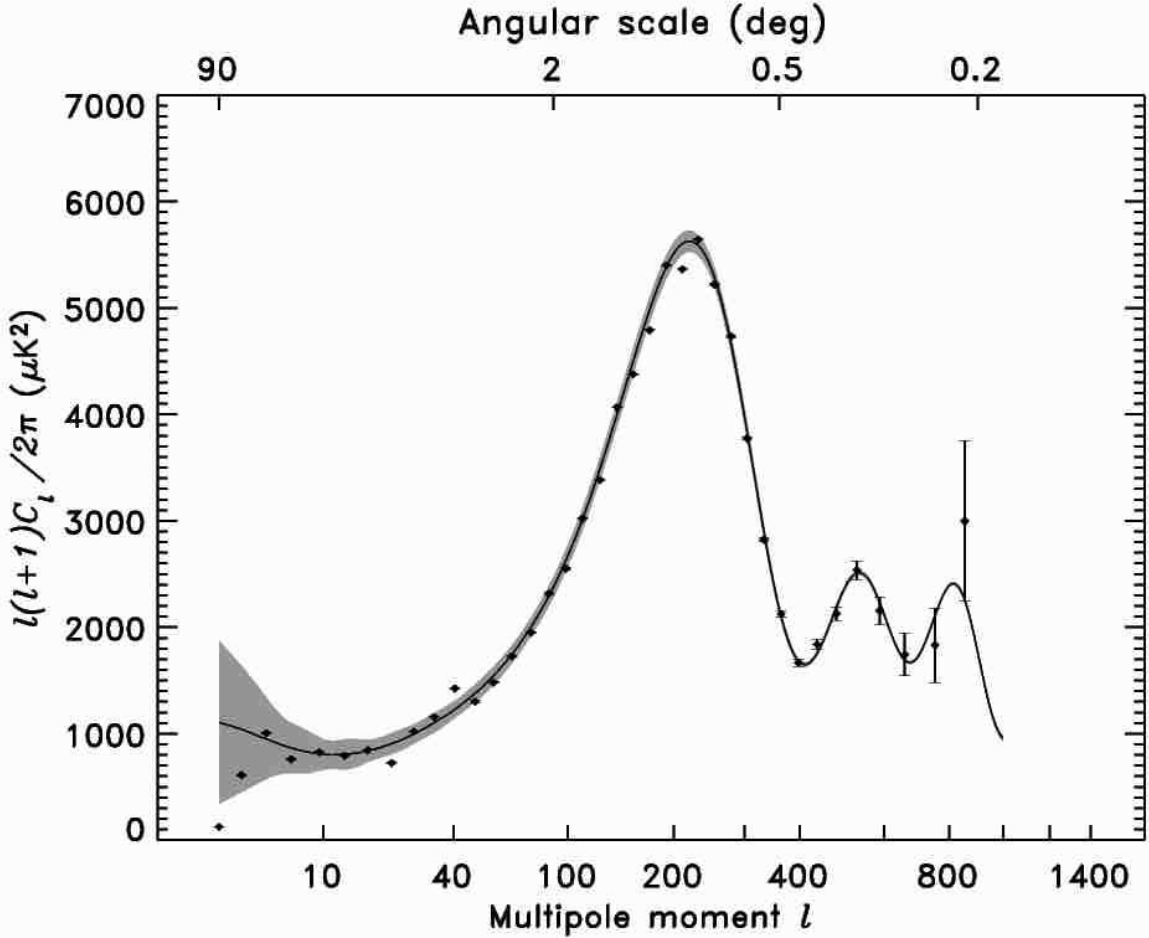}
\end{center}
\caption{The final angular power spectrum, $l(l+1)C_l/2\pi$, obtained  from the
28 cross-power spectra, as described in \S\ref{sec:combined}.  The data are
plotted with 1$\sigma$ measurement errors only which reflect the combined
uncertainty due to noise, beam, calibration, and source subtraction 
uncertainties.  The solid line shows the best-fit $\Lambda$CDM model from
\citet{spergel/etal:2003}.  The grey band around the model is the 1$\sigma$
uncertainty due to cosmic variance on the cut sky.  For this plot, both the 
model and the error band have been binned with the same boundaries as the data,
but they have been plotted as a splined curve to guide the eye. On the scale of
this plot the unbinned model curve would be virtually indistinguishable from
the binned curve except in the vicinity of the third peak.
\label{fig:final_spec}}
\end{figure}

Our best estimate of the angular power spectrum of the CMB is shown in 
Figure~\ref{fig:final_spec}.  Also shown is the best-fit $\Lambda$CDM model
from \citet{spergel/etal:2003} which is based on a fit to the this spectrum
plus  a compilation of additional CMB and large-scale structure data. The \map\
data points  are plotted with measurement errors based on the diagonal elements
of the Fisher matrix presented in Appendix~\ref{sec:fisher}.  The cosmic
variance errors, which include the effects of the sky cut, are plotted as a
1$\sigma$ band around the best-fit model.  As discussed in
\citet{spergel/etal:2003}, the model is an excellent fit to the data.  The
combined spectrum provides a definitive measurement of the CMB power spectrum
with uncertainties limited by cosmic variance up to $l \sim 350$.  The spectrum
clearly exhibits a first acoustic peak at $l=220.1 \pm 0.8$ and a second
acoustic  peak at $l=546 \pm 10$. \citet{page/etal:2003c} present an analysis
and  interpretation of the peaks and troughs in the first-year \map\ power
spectrum.

\begin{figure}
\begin{center}
\includegraphics[angle=90,width=1.0\textwidth]{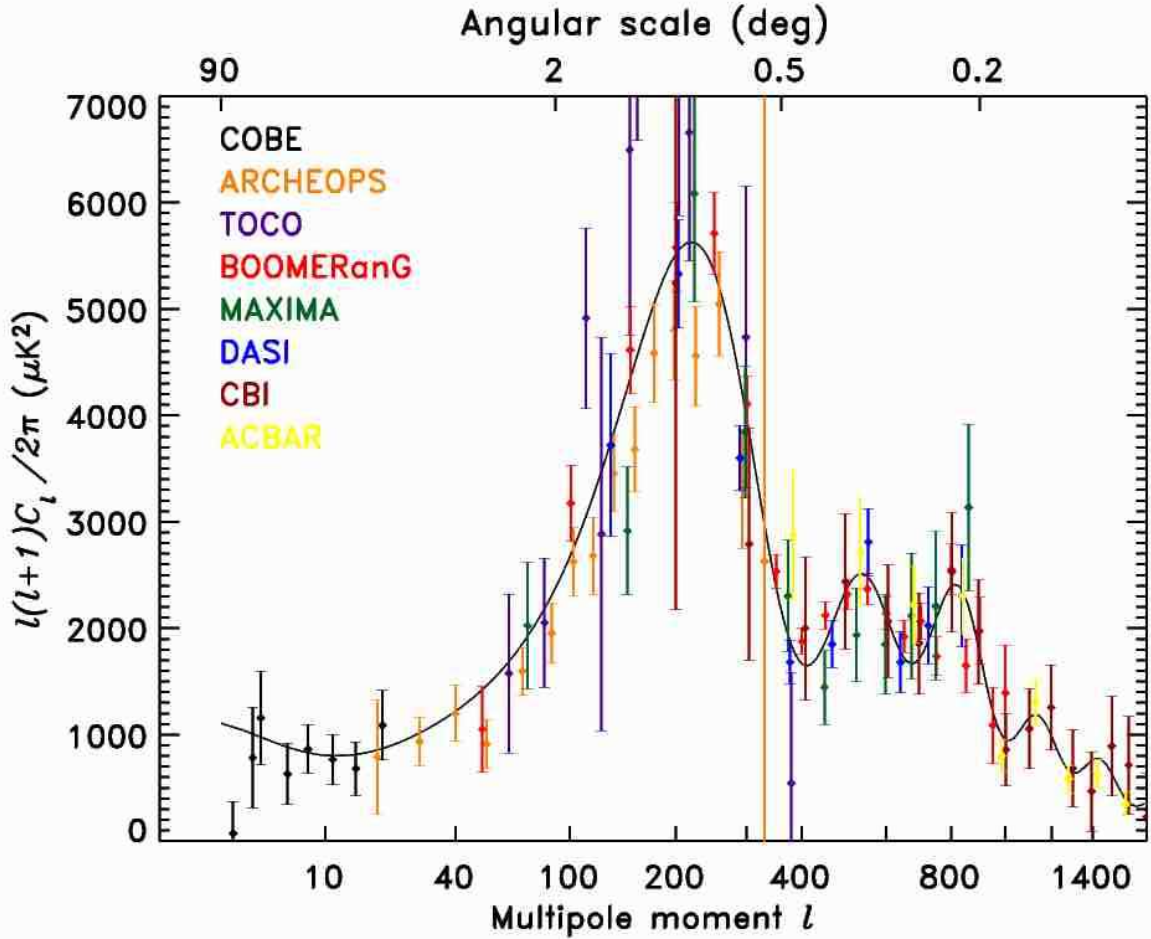}
\end{center}
\caption{A compilation of recent CMB power spectrum measurements compared to 
the best-fit $\Lambda$CDM model from the first-year \map\ data.  The data
points include noise and cosmic variance uncertainty (but not calibration
uncertainty) thus we omit the cosmic variance band from the model curve in the
Figure. On average, the pre-\map\ data agree well with the \map\ power
spectrum.
\label{fig:cf_comp}}
\end{figure}

Figure~\ref{fig:cf_comp} compares the first-year \map\ spectrum to a
compilation of recent balloon and ground-based measurements.  In order to make
this Figure meaningful, we plot the best-fit model spectrum to represent the
\map\ results.  The data points are plotted with errors that include both
measurement uncertainty and cosmic variance, so no error band is  included with
the model curve.  (Since individual groups report band power  measurement with
different band-widths, it is not possible to represent a single cosmic variance
band that applies to all data sets.)  The model spectrum fit to \map\ agrees
very well with the ensemble of previous observations.  

\begin{figure}
\begin{center}
\includegraphics[angle=90,width=1.0\textwidth]{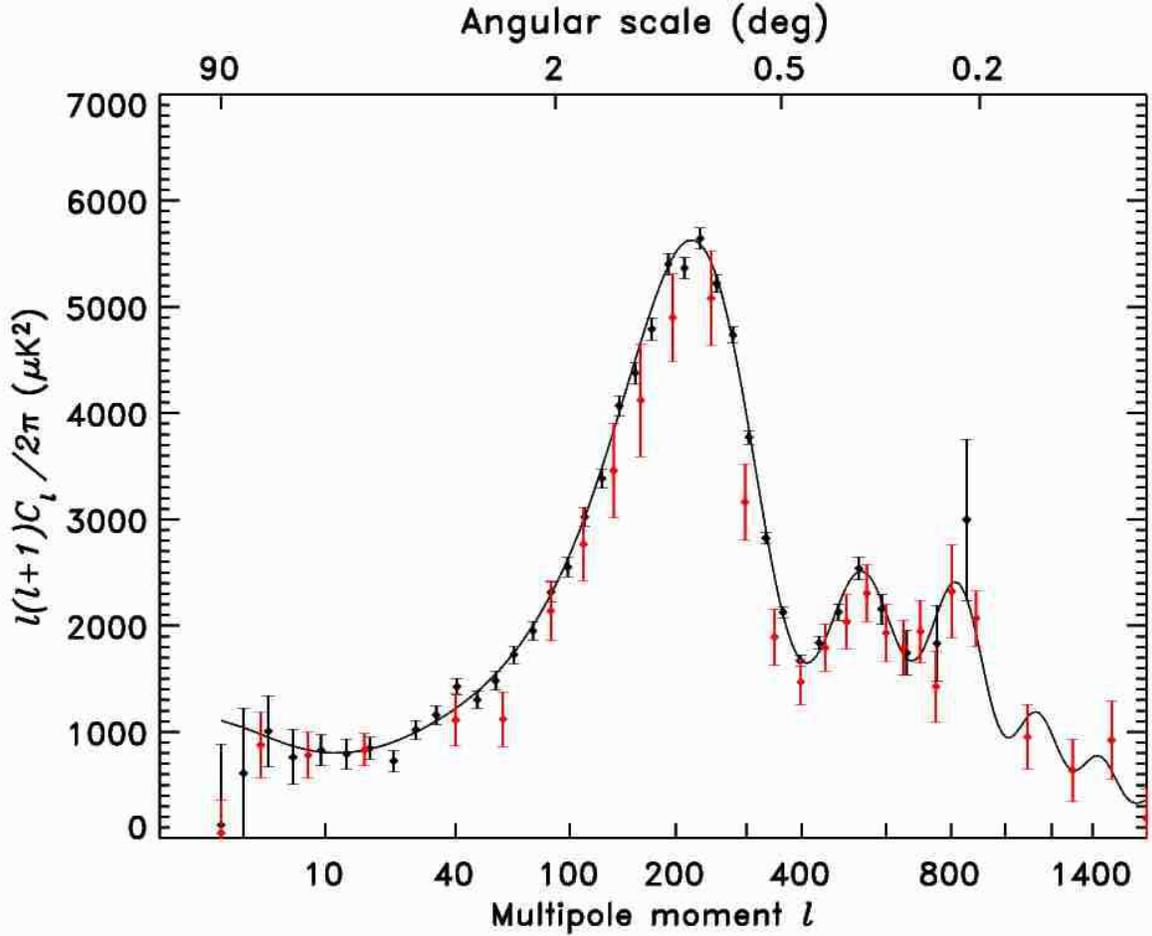}
\end{center}
\caption{The \map\ combined power spectrum, in black, compared to a compilation
of all  CMB data published prior to \map\ from \citet{wang/etal:2002}, in red. 
The \map\  data are plotted with cosmic variance plus measurement uncertainties
here in order to facilitate a comparison with the compiled data which is
reported in this way.  The data agree well on \cobe\ scales, $l < 20$, (but
note that the \map\ cosmic variance errors are computed from the best-fit model
rather than the data, thus they appear larger than the \cobe\ errors at the
quadrupole).  However, the  overall normalization of the \map\ spectrum is
$\sim$10\% higher  on smaller scales.
\label{fig:cf_wang}}
\end{figure}

\citet{wang/etal:2002} have  recently distilled a CMB power spectrum from an
optimal combination of  the extent pre-\map\ data.  In Figure~\ref{fig:cf_wang}
we plot their derived band power points alongside the \map\ data.  To make this
comparison meaningful, we plot the \map\ data with cosmic variance plus
measurement errors and omit the error band from the model spectrum.  The
distilled spectrum is notably lower than the \map\ data in the vicinity of the
first acoustic peak.  In a previous version of this work
\citep{wang/tegmark/zaldarriaga:2002} the authors noted  that the first
peak of their combined spectrum was lower than a significant fraction of their
input data.  They attribute this to  their formalism allowing for a
renormalization of individual experiments within their respective calibration
uncertainties.  Figure 1 in \citet{bennett/etal:2003} presents a similarly
distilled spectrum from the data extent in late 2001 and found a first peak
amplitude that was more intuitively consistent with the bulk of the input data,
and which is now seen to be consistent with the \map\ power spectrum.

\begin{figure}
\begin{center}
\includegraphics[angle=90,width=1.0\textwidth]{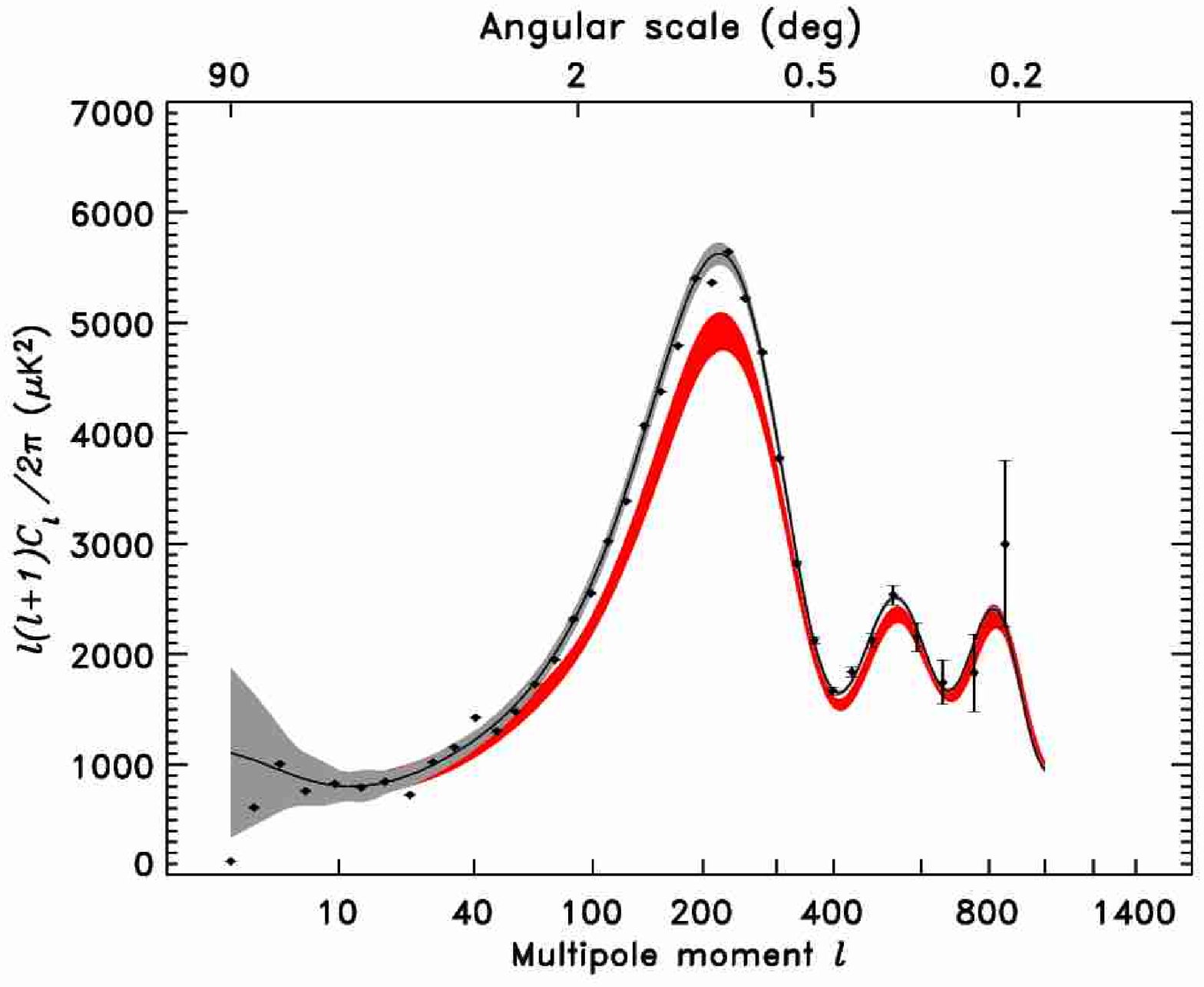}
\end{center}
\caption{The \map\ combined power spectrum compared to the locus of predicted
spectra, in red, based on a joint analysis of pre-\map\ CMB data and 2dFGRS 
large-scale structure data \citep{percival/etal:2002}.  As in
Figure~\ref{fig:final_spec}, the \map\ data are plotted with measurement
uncertainties, and the best-fit $\Lambda$CDM model \citep{spergel/etal:2003} is
plotted with a 1$\sigma$ cosmic variance error band.  The locus of predicted
spectra lie systematically below the \map\ data at intermediate $l$.
\label{fig:cf_2dFGRS}}
\end{figure}

Figure~\ref{fig:cf_2dFGRS} shows the \map\ combined power spectrum compared to
the locus of predicted spectra, in red, based on a joint analysis of pre-\map\
CMB data and 2dFGRS large-scale structure data \citep{percival/etal:2002}.  As
in Figure~\ref{fig:final_spec}, the \map\ data are plotted with measurement
uncertainties, and the best-fit $\Lambda$CDM model \citep{spergel/etal:2003} is
plotted with a 1$\sigma$ cosmic variance error band. 
\citet{percival/etal:2002} predict the location of the first peak should occur
at $l=221.8 \pm 2.4$, which is quite consistent with the value reported by
\citet{page/etal:2003c} of $l=220.1 \pm 0.8$. The height of the first peak was
predicted to be in the range $4920 \pm 170$ $\mu$K$^2$, while 
\citet{page/etal:2003c} report a measured height of $5580 \pm 75$ $\mu$K$^2$,
about 13\% higher.  Unlike the position, the amplitude of the first peak has a
complicated  dependence on cosmological parameters.  
\citet{percival/etal:2002} report best-fit parameters for a $\Lambda$CDM model
that are mostly consistent with those reported by \citet{spergel/etal:2003} for
the same class of models.  The only mildly disparate comparison lies in the
combination of normalization, $\sigma_8$, and optical depth, $\tau$.  
\citet{percival/etal:2002} report the product $\sigma_8 e^{-\tau} = 0.72
\pm 0.03 \pm 0.02$, where the first error is a ``theory'' error and the
second is measurement error. While \citet{spergel/etal:2003} does not
report a maximum likelihood range for this explicit parameter
combination, the product of their maximum likelihood values for
$\sigma_8$ and $\tau$ yields $\sigma_8 e^{-\tau} = 0.74$, which is
consistent with \citet{percival/etal:2002}, but would make the first
peak a few percent higher. Small differences in $\Omega_b h^2$,
$\Omega_m h^2$, and $n_s$, may also contribute to the difference.

\section{CONCLUSIONS}
\label{sec:conclude}

We present measurements of the angular power spectrum of the cosmic microwave
background from the first-year \map\ data.  The eight high-frequency sky maps
from DAs Q1 through W4 were used to estimate 28 cross-power spectra, which are
largely independent of the noise properties of the experiment.  These  data were
tested for consistency in \S\ref{sec:data}, then used in \S\ref{sec:combined}
as input to a final combined spectrum, discussed in \S\ref{sec:discuss}.  The
procedure for estimating the uncertainties in the final combined spectrum were
discussed in \S\ref{sec:covariance} and in numerous Appendices.

The combined spectrum provides a definitive measurement of the CMB power
spectrum, with uncertainties limited by cosmic variance up to $l \sim 350$,
and a signal to noise per mode $>1$ up to $l \sim 650$.  The spectrum clearly
exhibits a first acoustic peak at $l=220.1 \pm 0.8$ and a second acoustic  peak
at $l=546 \pm 10$. \citet{page/etal:2003c} present an analysis and
interpretation of the peaks and troughs in the first-year \map\ power
spectrum.  \citet{spergel/etal:2003}, \citet{verde/etal:2003}, and
\citet{peiris/etal:2003} analyze the combined spectrum in the context of
cosmological models. They conclude that the data provide strong support
adiabatic initial conditions, and they give precise measurements of a number of
cosmological parameters.   \citet{kogut/etal:2003} analyze the correlation
between \map's temperature and polarization signals, the $C_l^{TE}$ spectrum,
and present evidence for a relatively high optical depth, and an early period
of cosmic reionization. Among other things, this result implies that the
temperature power spectrum is suppressed by $\sim$30\% on degree angular
scales, due to secondary scattering.

A variety of first-year \map\ data products are being made available by NASA's 
new Legacy Archive for Microwave Background Data Analysis (LAMBDA).  In
addition to the sky maps and calibrated time-ordered data, we are providing the
28  cross power spectra used in this paper (with diffuse foregrounds
subtracted), the combined spectrum from \S\ref{sec:opt_combined}, and a Fortran
90 subroutine to compute the likelihood of a given cosmological model, (the
code will also  optionally return the Fisher (inverse-covariance) matrix for
the combined spectrum.) The LAMBDA URL is \verb"http://lambda.gsfc.nasa.gov/".

\acknowledgements

The \map\ mission is made possible by the support of the Office of Space 
Sciences at NASA Headquarters and by the hard and capable work of scores of 
scientists, engineers, technicians, machinists, data analysts, budget
analysts,  managers, administrative staff, and reviewers.  We thank Mike Nolta
for helpful comments on an earlier draft of this manuscript.  LV is supported
by NASA through Chandra Fellowship PF2-30022 issued by the Chandra X-ray
Observatory center, which is operated by the Smithsonian Astrophysical
Observatory for an on behalf of  NASA under contract NAS8-39073. We acknowledge
use of the HEALPix package.

\appendix

\section{POWER SPECTRUM ESTIMATION METHODS}
\label{sec:ps_estimation}

For the analysis of \map's first-year data, we have chosen two distinct methods
for  inferring the power spectrum.  The first is a fast and accurate quadratic 
method for estimating the power spectrum of a partial sky map 
\citep{hivon/etal:2002}.  We summarize the basic approach here, highlighting 
the aspects of the method that are especially pertinent to \map, and refer  the
reader to \citet{hivon/etal:2002} for details.  The second is a maximum
likelihood method that provides an independent estimate of the spectrum
measured by \map\ \citep{oh/spergel/hinshaw:1999}.  

\begin{figure}
\begin{center}
\includegraphics[angle=90,width=1.0\textwidth]{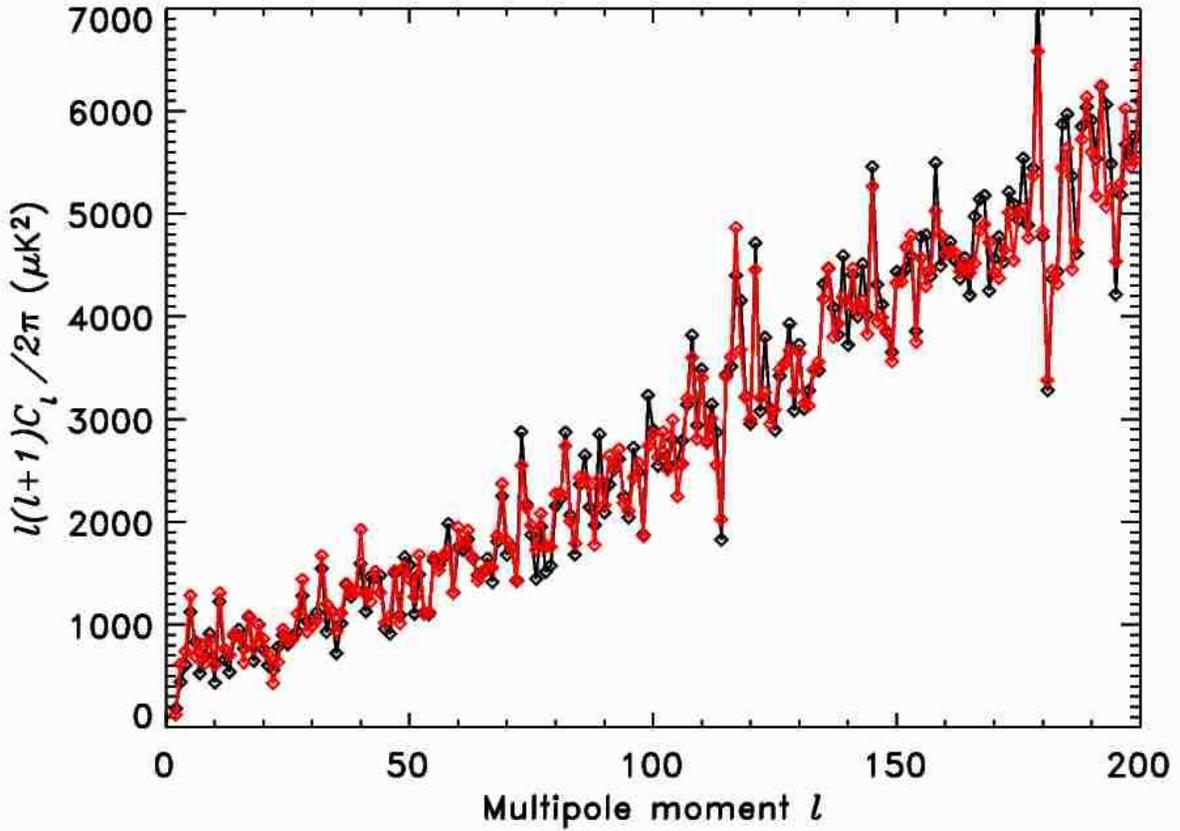}
\end{center}
\caption{A comparison of two power spectra computed from a single V band map.
The black points result from the maximum likelihood method, the red points from
the quadratic estimator computed with uniform pixel weighting.  In both cases
a noise model has been assumed to treat the noise bias.  At higher $l$,
the two spectra would be nearly identical since both impose inverse noise
weighting on the data.
\label{fig:cf_method}}
\end{figure}

We have applied both of these methods to the \map\ data. The results are shown
in Figure~\ref{fig:cf_method}, which shows spectra estimated from the V band
map, up to $l=200$, for the two methods.  The maximum likelihood estimate has
slightly lower uncertainties at low $l$, because the method optimally weights
the data with a pixel-pixel covariance ${\bf C = S+N \approx S}$ where ${\bf S}$
is the covariance of the CMB signal and ${\bf N}$ is the covariance of the
noise (see Appendix~\ref{sec:max_like}).  Our quadratic estimator uses uniform
pixel weights at low $l$ (see Appendix~\ref{sec:weights}) though it is clear
from  the Figure that the difference is not significant. At high $l$,
where the \map\ data are noise dominated, the two estimators give essentially
identical results because they effectively weight the data in the same way.

To obtain our ``best''  estimate of the \map\ power spectrum, we adopt the
quadratic estimator  because it can be readily applied to pairs of \map\
radiometers in a way  that is nearly independent of the properties of the
instrument noise.  In \S\ref{sec:combined} we discuss our methodology for
combining spectra from  pairs of radiometers and present the final combined
spectrum.

\subsection{Quadratic Estimation}
\label{sec:quad_estimation}

\citet{hivon/etal:2002} start with the full-sky estimator in
equation~(\ref{eq:int_alm_full}), add a position dependent weight, $W(\n)$, and
set $W$ to zero in the regions where the sky is  contaminated. In other
regions, $W$ is chosen to optimize the sensitivity of the  estimator. A
temperature map $\Delta T(\n)$ on which a weight $W(\n)$ is applied can be
decomposed in spherical harmonic coefficients as
\begin{eqnarray}
\tilde{a}_{l m} & = & \int d\Omega_{\n} \, \Delta T(\n)\,W(\n)\,Y^*_{l m}(\n)
\label{eq:alm_cutsky_sum} \\
   & \approx & \Omega_p \sum_p \Delta T(p)\,W(p)\,Y^*_{l m}(p), 
\label{eq:alm_cutsky}
\end{eqnarray}
where the integral over the sky is approximated by a discrete sum over map
pixels, each of which subtends solid angle $\Omega_p$.  
\citet{hivon/etal:2002} then define the ``pseudo power spectrum'' 
$\tilde{C}_l$ as
\begin{equation}
\tilde{C}_l = \frac{1}{2l+1} \sum_{m=-l}^{l}
  \left|\tilde{a}_{l m}\right|^2.
\label{eq:Cl_cutsky}
\end{equation}

The pseudo power spectrum $\tilde{C}_l$, given by the weighted spherical 
harmonic transform of a map, is clearly different from the full sky angular 
spectrum, $C^{\rm sky}_l$, but the {\em ensemble averages} of the two spectra 
can be related by 
\begin{equation}
\VEV{\tilde{C}_l} = \sum_{l'} G_{ll'}\,\VEV{C^{\rm sky}_{l'}}
\label{eq:model_cutsky}
\end{equation}
where $G_{ll'}$ describes the mode coupling resulting from the 
weight function $W(\n)$ \citep{hauser/peebles:1973}.  \citet{hivon/etal:2002}
give the following expression for the coupling matrix, which depends only
on the geometry of the weight function $W(\n)$
\begin{equation}
G_{l_1l_2} = \frac{2l_2+1}{4\pi} \sum_{l_3}
  \,(2l_3+1)\,{\mathcal{W}}_{l_3}\,\wjjj{l_1}{l_2}{l_3}{0}{0}{0}^2,
\label{eq:kernel_final}
\end{equation}
where the final term in parentheses is the Wigner 3-$j$ symbol, and 
$\mathcal{W}_l$ is the angular power spectrum of the weight function
\begin{equation}
\mathcal{W}_l = \frac{1}{2l+1}\sum_{m}\,\left|w_{l m}\right|^2,
\end{equation}
where
\begin{equation}
w_{l m} = \int d\Omega_{\n} \,W(\n)\,Y_{l m}^*(\n).
\label{eq:power_window}
\end{equation}

Upon inverting the coupling matrix $G_{ll'}$ and making the identification
$\VEV{C^{\rm sky}_l} = C_l$, we obtain the following estimator of the power 
spectrum
\begin{equation}
C^{\rm obs}_l = \sum_{l'} G^{-1}_{ll'}\,\tilde{C}_{l'}.
\label{eq:ps_cutsky}
\end{equation}
The computation of equation~(\ref{eq:alm_cutsky}) for each $(l,m)$ up to
$l=\lmax$ would scale as $N_{\rm pix}\lmax^2$ if performed on an arbitrary 
pixelization of the sphere, where $N_{\rm pix}$ is the number of sky map
pixels. However, for a pixelization scheme with iso-latitude pixel centers,
fast FFT methods may be employed to speed up the evaluation, so it scales like
$N_{\rm pix}^{1/2} \lmax^2$ \citep{muciaccia/natoli/vittorio:1997}.  The \map\
sky maps have been produced using the HEALPix layout
\citep{gorski/hivon/wandelt:1998} which supports such fast  spherical harmonic
transforms.  In particular, the HEALPix routine \verb"map2alm"\ evaluates
equation~(\ref{eq:alm_cutsky}).

\subsubsection{Auto- and Cross-Power Spectra from the \map\ Data}
\label{sec:cross_def}

For a multi-channel experiment like \map\ it is quite powerful to evaluate 
the power spectra from different maps and compare results.  In particular, the 
quadratic estimator described above may be used on 1 or 2 maps at a time by
generalizing the expression for the pseudo power spectrum
equation~(\ref{eq:Cl_cutsky})  as
\begin{equation}
\tilde{C}_l = \frac{1}{2l+1} \sum_{m=-l}^{l}
  \tilde{a}^i_{l m} \tilde{a}^{j\ast}_{l m}
\label{eq:cl_crosssky}
\end{equation}
where $\tilde{a}^i_{l m}$ refers to the transform of map $i$ and 
$\tilde{a}^{j\ast}_{l m}$ refers to map $j$, which needn't be the same as  map
$i$.  As discussed in \S\ref{sec:instrument}, if $i \ne j$ and the noise in
the two maps is uncorrelated, the estimator equation~(\ref{eq:cl_crosssky})
provides an unbiased estimate of the underlying power spectrum.  

\begin{figure}
\begin{center}
\includegraphics[width=0.8\textwidth]{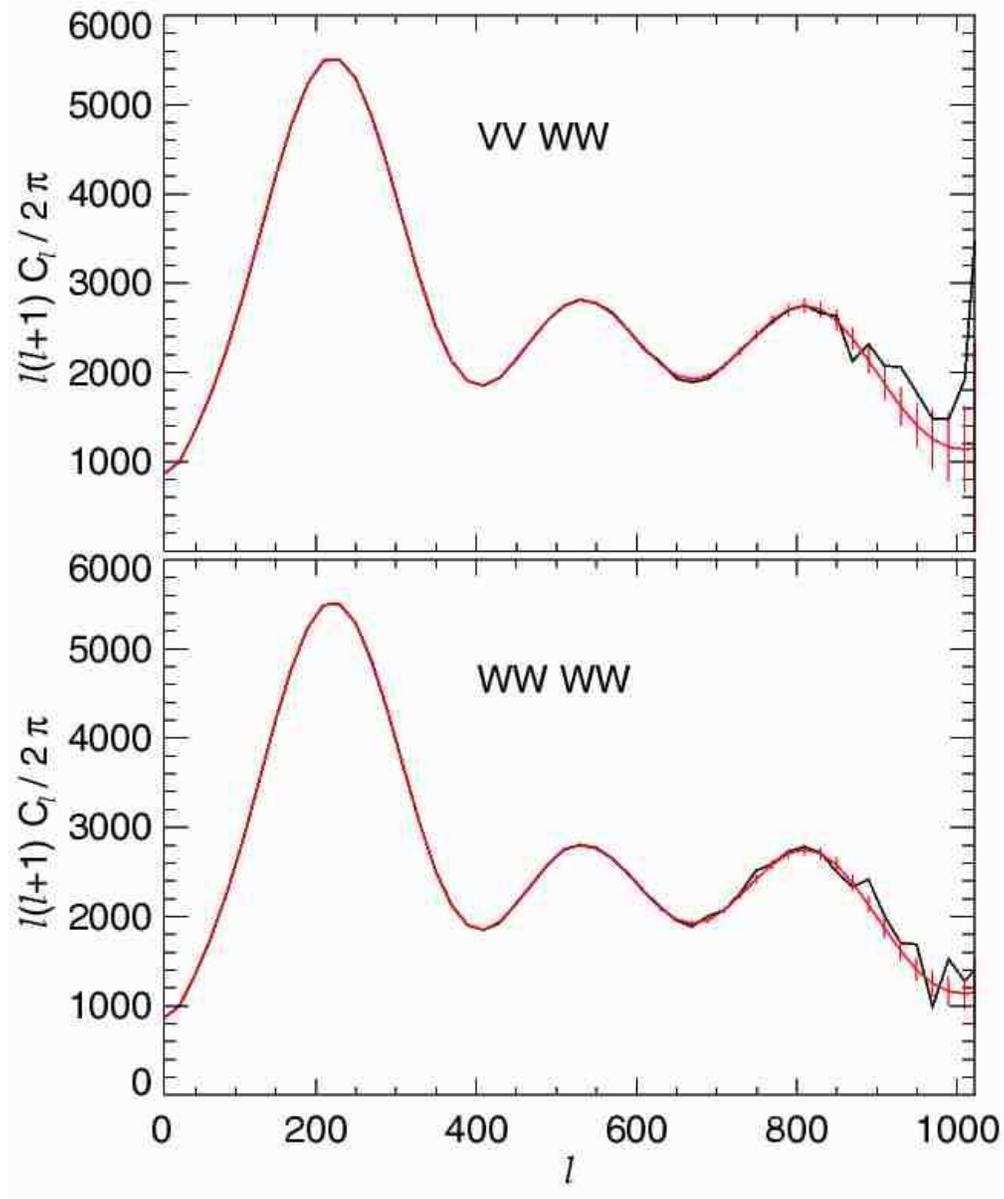}
\end{center}
\caption{The quadratic estimator discussed in \S\ref{sec:quad_estimation} has
been extensively tested with Monte Carlo simulations of first-year \map\
data.   The top panel shows a model spectrum in red, and the mean of 600
realizations  of the cross-power spectrum, computed from the V and W
band-averaged maps, in  black.  The red error bars show the uncertainty in the
mean in bins with $\Delta l = 20$.  The cross-power spectrum estimator is
unbiased.  The  bottom panel is similar, but with auto-power spectra computed
from the W  band-averaged map.  Here the noise bias term was estimated directly
from the high  $l$ tail of the computed spectrum. This simple auto-power
spectrum estimator is unbiased if the noise is white, and is  suitable for V
band data, but see Figure~\ref{fig:noise_spec} for examples of W band noise
spectra.
\label{fig:quad_bias_mc}}
\end{figure}

We have tested the auto- and cross-power estimator extensively with Monte Carlo
simulations of the first-year \map\ data.  Selected results from this testing 
are shown in Figure~\ref{fig:quad_bias_mc}.  The auto- and cross-power spectra 
that obtain from the flight data are presented in detail in \S\ref{sec:data}. 
The cross-power spectra form the basis for our final combined spectrum, 
presented in \S\ref{sec:combined}.

\subsubsection{Choice of Weighting}
\label{sec:weights}

We seek a weighting scheme that mimics the maximum likelihood estimation
(Appendix~\ref{sec:max_like}), which effectively weights the data by the full
inverse covariance matrix  ${\bf C}^{-1} = ({\bf S + N})^{-1}$.  For the
combined spectrum presented in \S\ref{sec:combined}, we use three distinct 
weighting functions in three separate $l$ ranges.
\begin{enumerate}
\item For $l < 200$ we give equal weight to all un-cut pixels,
\begin{equation}
W(p) = M(p)
\end{equation}
where $M(p)$ is the Kp2 sky mask, defined by \citet{bennett/etal:2003c}.  It
takes values of 0 within the mask and 1 otherwise.
\item For $l>450$ we use inverse-noise weighting,
\begin{equation}
W(p) = M(p) N_{\rm obs}(p)
\end{equation}
where $N_{\rm obs}(p)$ is the number of observations of pixel $p$.
\item For $200<l<450$ we use a transitional weighting,
\begin{equation}
W(p) = \frac{M(p)}{1/\VEV{N_{\rm obs}} + 1/N_{\rm obs}(p)}
\end{equation}
where $\VEV{N_{\rm obs}}$ is the mean of $N_{\rm obs}$ evaluated over the cut
sky.
\end{enumerate}
\citet{verde/etal:2003} discuss the choice of weighting in more detail.

\subsection{Maximum Likelihood Estimation}
\label{sec:max_like}

This Appendix provides a summary of the maximum likelihood approach to power
spectrum estimation originally presented by \citet{oh/spergel/hinshaw:1999}. 
If the temperature fluctuations are Gaussian, and the {\it a priori} 
probability of a given set of cosmological parameters is uniform, then the 
power spectrum may be estimated by maximizing the multi-variate Gaussian 
likelihood function
\begin{equation}
{\cal L}(C_l|{\bf m}) = \frac{\exp(-\frac{1}{2}
{\bf m}^T \, {\bf C}^{-1} \, {\bf m})}
{\sqrt{\det {\bf C}}}
\end{equation}
where ${\bf m}$ is a data vector (see below) and ${\bf C}$ is the covariance
matrix of the data, which has contributions from both the signal and the
instrument noise,  ${\bf C} = {\bf S} + {\bf N}$.  We can work in whatever
basis is most convenient. In the pixel basis the data are the sky map pixel
temperatures, and in the spherical harmonic basis the data are the 
$a_{l m}$ coefficients of the map.  In the former basis the noise 
covariance is nearly diagonal, while in the latter, the signal covariance is.
$$
\begin{array}{lllp{5mm}lll}
 & & \mbox{Pixel basis:} & & & & \mbox{Spherical harmonic basis:} \\[2mm]
{\bf m} & \rightarrow & T_i & & {\bf m} & \rightarrow & a_{l m} \\[2mm]
{\bf S} & \rightarrow & \sum_l \frac{(2l + 1)}{4\pi}C_lP_l(\cos\theta_{ij}) 
& & {\bf S} & \rightarrow & {\rm diag}(C_2,C_2,\ldots C_3,C_3,\ldots) \\[2mm]
{\bf N} & \rightarrow & \sigma_i^2 \delta_{ij} & & {\bf N} & \rightarrow & 
N_{(l m)(l m)'} \mbox{(see below)}.
\end{array}
$$
For \map, the length of the data vector, $N_{\rm data}$, is 2,672,361, the
number of $7'$ sky map pixels (HEALPix $N_{\rm side}=512$) that survive the
Galaxy cut, so it is  necessary to find methods for evaluating ${\cal L}$ that
do not require a  full inversion of the covariance matrix ${\bf C}$, which
requires   $O\left(N_{\rm data}^3\right)$ operations.  We use an iterative
method for  evaluating the  likelihood that exploits the ability to  find an
approximate inverse $\tilde{\bf C}^{-1}$.  The most important features in the
data that make this possible are that \map\ observes the full sky and the
Galaxy cut is predominantly azimuthally  symmetric in Galactic coordinates
\citep{bennett/etal:2003b}.  Of secondary importance for this pre-conditioner
is that \map's noise per pixel does not vary strongly across the sky
\citep{bennett/etal:2003}.  We discuss the pre-conditioner in more detail 
below.

Defining $f \equiv -2\ln {\cal L}$ and 
${\bf P}_l \equiv \frac{\partial {\bf C}}{\partial C_l}$, we maximize the 
likelihood by solving
\begin{equation}
\frac{\partial f}{\partial C_l} = 0
 = {\bf m}^T\,{\bf C}^{-1}\,{\bf P}_l\,{\bf C}^{-1}\,{\bf m} 
 + {\rm tr}({\bf C}^{-1}\,{\bf P}_l)
\end{equation}
using a Newton-Raphson root finding method that generates an iterative
estimate of the angular power spectrum
\begin{equation}
C_l^{(n+1)} = C_l^{(n)} - \frac{1}{2}\sum_{l'} 
F_{ll'} \frac{\partial f}{\partial C_{l'}}
\label{eq:nr_cl}
\end{equation}
at each step.  Here $F_{ll'}$ is the Fisher matrix
\begin{equation}
F_{ll'} = -\left<\partial^2 \ln {\cal L} /
          \partial C_l\partial C_{l'} \right>
 = \frac{1}{2}{\rm tr}({\bf C}^{-1}\,{\bf P}_l\,{\bf C}^{-1}\,{\bf P}_{l'}).
\label{eq:fisher}
\end{equation}
To implement the solution in equation~(\ref{eq:nr_cl}) we need a fast way to 
evaluate the following three components of $\sum_{l'} F_{ll'}
\frac{\partial f}{\partial C_{l'}}$:
\begin{enumerate}
\item ${\bf m}^T\,{\bf C}^{-1}\,{\bf P}_l\,{\bf C}^{-1}\,{\bf m}$
\item ${\rm tr}({\bf C}^{-1}\,{\bf P}_l)$
\item ${\rm tr}({\bf C}^{-1}\,{\bf P}_l\,{\bf C}^{-1}\,{\bf P}_{l'})$.
\end{enumerate}
We use the spherical harmonic basis in which the data vector consists of the 
$a_{l m}$ coefficients of the map obtained by least squares fitting on the 
cut sky.  The signal covariance is diagonal in this basis, while the noise 
matrix is obtained from the normal equations for the
$a_{l m}$ fit
\begin{equation}
\sum_{(l m)'} N^{-1}_{(l m)(l m)'} a_{(l m)'}
 =  y_{(l m)}
\end{equation}
where
\begin{eqnarray}
N^{-1}_{(l m)(l m)'} & \equiv & 
\sum_i \frac{Y_{(l m)}(\hat{n}_i)Y_{(l m)'}(\hat{n}_i)}{\sigma_i^2} \\
y_{(l m)} & \equiv &
\sum_i \frac{T_i Y_{(l m)}(\hat{n}_i)}{\sigma_i^2}.
\label{eq:ny_def}
\end{eqnarray}
The sums are over all sky map pixels that survive the Galaxy cut, and we have
assumed that  the noise is uncorrelated from pixel to pixel.

\subsubsection{Evaluation of ${\bf C}^{-1}\,{\bf m}$}

The term ${\bf C}^{-1}\,{\bf m}$ appears repeatedly in the evaluation of 
equation~(\ref{eq:nr_cl}).  We compute this by solving ${\bf C}\,{\bf z} = 
({\bf S}+{\bf N})\,{\bf z} = {\bf m}$ for ${\bf z}$.  A more numerically 
tractable system is obtained by multiplying both sides by
${\bf S}^{\frac{1}{2}}\,{\bf N}^{-1}$, so
\begin{equation}
({\bf I} + {\bf S}^{\frac{1}{2}}\,{\bf N}^{-1}\,{\bf S}^{\frac{1}{2}})
\,{\bf S}^{\frac{1}{2}}\,{\bf z}
 = {\bf S}^{\frac{1}{2}}\,{\bf N}^{-1}\,{\bf m}
 = {\bf S}^{\frac{1}{2}}\,{\bf y}
\label{eq:czm}
\end{equation}
where ${\bf y}$ is the spherical harmonic transform of the map, defined in 
equation~(\ref{eq:ny_def}).  Note that ${\bf y}$ can be quickly computed in
any  pixelization scheme that has iso-latitude pixel centers with fixed
longitude spacing.  We then solve equation~(\ref{eq:czm}) using an iterative
conjugate  gradient method with a pre-conditioner for the matrix  ${\bf A}
\equiv ({\bf I} + {\bf S}^{\frac{1}{2}}\,{\bf N}^{-1}\, {\bf
S}^{\frac{1}{2}})$.  We find the following block-diagonal form of ${\bf A}$ to
be a good starting point
\begin{equation}
\tilde{{\bf A}} = \left( \begin{array}{cc}
{\bf I} + {\bf S}^{\frac{1}{2}}\,\tilde{{\bf N}}^{-1}\,{\bf S}^{\frac{1}{2}} & 0 \\
0 & \mbox{diag}({\bf I} + {\bf S}^{\frac{1}{2}}\,\tilde{{\bf N}}^{-1}\,{\bf S}^{\frac{1}{2}})
\end{array} \right)
\label{pre_con}
\end{equation}
where $\tilde{{\bf N}}^{-1}$ is a block-diagonal approximation of the
noise  matrix discussed below.  The lower-right block of $\tilde{{\bf A}}$
occupies  the high $l$ portion of the matrix where the signal to noise ratio 
${\bf S}^{\frac{1}{2}}\,{\bf N}^{-1}\,{\bf S}^{\frac{1}{2}}$ is low,  so a
diagonal approximation is adequate.  The upper-left block occupies the  low $l$
portion of the matrix where the signal dominates the noise, so we  need a
better estimate of ${\bf N}^{-1}$.  In practice we find this split works  well
at $l = 512$ for the \map\ noise levels.  As for the  approximate form of
${\bf N}^{-1}$, defined in equation~(\ref{eq:ny_def}), note that the  dominant
off-diagonal contributions arise from the azimuthally symmetric Galaxy  cut,
which couples different $l$ modes, but not $m$ modes.  Thus  ${\bf N}^{-1}$ is
approximately block diagonal, with perturbations induced by  the non-uniform
sky coverage of \map.  We therefore use a block diagonal form  of ${\bf
N}^{-1}$ as the pre-conditioner,
\begin{equation}
\tilde{N}^{-1}_{(l m)(l m)'} = N^{-1}_{(l m)(lm)'}\delta_{mm'}.
\end{equation}
Using the pre-conditioner equation~(\ref{pre_con}) we find that the conjugate
gradient solution of equation~(\ref{eq:czm}) converges in approximately six
iterations and requires only cpu-minutes of processing on an SGI Origin
2000.

\section{POINT SOURCE SUBTRACTION}
\label{sec:marginalize}

In this Appendix we describe the procedure we use to estimate and subtract the
point source contribution directly from the multi-frequency cross-power
spectra.  We then show how we incorporate the source model uncertainty 
into the covariance matrix of the source-subtracted spectra by marginalizing
a Gaussian likelihood function over the source model amplitude parameter.

This marks the starting point of the multi-frequency analysis which will lead 
to the combined power spectrum, discussed in \S\ref{sec:opt_combined}. In
order  to generate the combined spectrum, we need the full covariance matrix of
the cross-power spectra (see \S\ref{sec:covariance}).  Our approach to
generating the full covariance is to start with the ideal, full-sky form,
which only includes cosmic variance and instrument noise, then we incorporate
additional effects step by step, as outlined in \S\ref{sec:opt_combined} and
in these Appendices.  For an ideal experiment with full sky coverage, no point 
source contamination, and no beam uncertainty the covariance matrix is
\begin{equation}
\Sigma^{ii' jj'}_{ll} = \frac{1}{(2l+1)}
\left[\left(C^{\rm th}_lw^{ij}_l+n^{i}n^{j}\delta_{ij}\right)
      \left(C^{\rm th}_lw^{i'j'}_l+n^{i'}n^{j'}\delta_{i'j'}\right)
     +\left(C^{\rm th}_lw^{ij'}_l+n^{i}n^{j'}\delta_{ij'}\right)
      \left(C^{\rm th}_lw^{i'j}_l+n^{i'}n^{j}\delta_{i'j}\right)\right]
\label{eq:cl_cov_diag}
\end{equation}
where $\delta_{ij}$ denotes the Kronecker symbol, $w^{ij}_l \equiv 
b^i_l\,b^j_l\,p^2_l$ is the window function, and $n^i n^{i'}\delta_{ii'} = N^{\bf i}$.

\subsection{Estimating the Point Source Amplitude}
\label{sec:amp_ps}

We start by assuming a Gaussian likelihood for the sky model, given the \map\
data
\begin{equation}
-2\ln{\cal L}(A,C_l\,\vert\,C^{\bf i}_l) = \sum_{{\bf ij}\,\,l}\,\,
 [\,C^{\bf i}_l - (C_l+AS^{\bf i})w^{\bf i}_l\,]
 \,\,(\Sigma^{-1})^{\bf ij}_l\,\,
 [\,C^{\bf j}_l - (C_l+AS^{\bf j})w^{\bf j}_l\,],
\label{eq:src_like_diag}
\end{equation}
where $C^{\bf i}_l$ is cross-power spectrum ${\bf i}$, $w^{\bf i}_l$ is the
window function for spectrum ${\bf i}$, $C_l$ is the true CMB power spectrum,
and  $AS^{\bf i}$ is the source model defined in
equations~(\ref{eq:source_model}) and (\ref{eq:spec_model}). Here we assume
the  diagonal form of the covariance matrix in
equation~(\ref{eq:cl_cov_diag}). 

To determine the best-fit source amplitude, $A$, we marginalize this likelihood
over the CMB spectrum, $C_l$. First we expand 
equation~(\ref{eq:src_like_diag}) as
\begin{eqnarray}
-2\ln{\cal L} & = & \sum_l \sum_{\bf ij} \,\,
(C^{\bf i}_l - AS^{\bf i}w^{\bf i}_l\,) (\Sigma^{-1})^{\bf ij}_l (C^{\bf j}_l-AS^{\bf j} w^{\bf j}_l\,) \nonumber \\
 & - & \sum_l 2C_l \sum_{\bf ij} \,\,
(C^{\bf i}_l - AS^{\bf i}w^{\bf i}_l\,) (\Sigma^{-1})^{\bf ij}_l w^{\bf j}_l\, \nonumber \\
 & + & \sum_l C^2_l \sum_{\bf ij} \,\, 
w^{\bf i}_l\,(\Sigma^{-1})^{\bf ij}_l w^{\bf j}_l\,,
\end{eqnarray}
which is a quadratic form in $C_l$, ${\cal L} \propto \prod_l 
\exp[-\case{1}{2}(aC_l^2-2bC_l+c)]$, with
\begin{eqnarray}
a & \equiv & \sum_{\bf ij} \,\,
  w^{\bf i}_l\,(\Sigma^{-1})^{\bf ij}_l \,w^{\bf j}_l, \nonumber \\
b & \equiv & \sum_{\bf ij} \,\,
  (C^{\bf i}_l - AS^{\bf i}w^{\bf i}_l\,) (\Sigma^{-1})^{\bf ij}_l w^{\bf j}_l, \nonumber \\
c & \equiv & \sum_{\bf ij} \,\,
  (C^{\bf i}_l - AS^{\bf i}w^{\bf i}_l\,) (\Sigma^{-1})^{\bf ij}_l (C^{\bf j}_l-AS^{\bf j} w^{\bf j}_l\,).
\end{eqnarray}
We wish to evaluate the marginalized likelihood, 
${\cal L}_{C_l}(A) = \int dC_l {\cal L}(A,C_l)$. 
This is readily evaluated using the substitution $C_l \to C_l - b/a$, giving
${\cal L}_{C_l} \propto \prod_l \exp[-\case{1}{2}(c-b^2/a)]$, where we drop a 
multiplicative term proportional to $a$, which is independent of $A$. 
The marginalized likelihood function is thus
\begin{eqnarray}
-2\ln{\cal L}_{C_l} & = & \sum_l \sum_{\bf ij}
(C^{\bf i}_l-AS^{\bf i}w^{\bf i}_l)(\Sigma^{-1})^{\bf ij}_l(C^{\bf j}_l-AS^{\bf j}w^{\bf j}_l) \nonumber \\
 & - & \sum_l \frac{1}{\sum_{\bf i'j'}w^{\bf i'}_l\,(\Sigma^{-1})^{\bf i'j'}_l \,w^{\bf j'}_l}
\left[\sum_{\bf ij}\,(C^{\bf i}_l - AS^{\bf i}w^{\bf i}_l\,)(\Sigma^{-1})^{\bf ij}_l w^{\bf j}_l\right]^2.
\end{eqnarray}
Setting $\partial{\cal L}_{C_l}(A)/\partial A = 0$ gives the most likely point 
sources amplitude
\begin{equation}
\bar{A} = \frac{\sum_{{\bf ij}\,\,l}\,C^{\bf j}_l\,(\Sigma^{-1})^{\bf ij}_l\,H^{\bf i}_lw_l^{\bf i}}
               {\sum_{{\bf ij}\,\,l}\,S^{\bf j}w_l^{\bf j}\,(\Sigma^{-1})^{\bf ij}_l\,H^{\bf i}_l w_l^{\bf i}},
\label{eq:ps_amplitude}
\end{equation}
where
\begin{equation}
H^{\bf i}_{ l} = S^{\bf i} - \frac{\sum_{\bf i'j'}\,S^{\bf i'}\,(\Sigma^{-1})^{\bf i'j'}_l w_l^{\bf j'}}
                                  {\sum_{\bf i'j'}\,w_l^{\bf i'}(\Sigma^{-1})^{\bf i'j'}_lw_l^{\bf j'}}.
\end{equation}
The standard error on the best fit value of $A$ is
\begin{equation}
\sigma^2_{\rm src} = \left[\sum_{{\bf ij}\,\,l}\,H^{\bf i}_l
 \, w_l^{\bf i}(\Sigma^{-1})^{\bf ij}_l\,S^{\bf j}_l w_l^{\bf j}\right]^{-1}.
\end{equation}
For \map, the off-diagonal terms in the covariance matrix are small. Here, neglecting
them changes the inferred point source amplitude by less than 1\%.

\subsection{Marginalizing over Point Source Amplitude}
\label{sec:marg_ps}

The source subtraction procedure discussed above is uncertain.  In this section
we incorporate this uncertainty into the full covariance matrix for the 
cross-power spectra.  We again assume a Gaussian likelihood function of the
form
\begin{equation}
-2\ln{\cal L} = \sum_{{\bf ij}\,\,ll'} \,\, 
[\bar{C}^{\bf i}_l - (C^{\rm th}_l + \delta A S^{{\bf i}}_l)w^{\bf i}_l\,]
\,\,(\Sigma^{-1})^{\bf ij}_{ll'}\,\,
[\bar{C}^{\bf j}_{l'} - (C^{\rm th}_{l'} + \delta A S^{{\bf j}}_{l'})w^{\bf j}_{l'}\,]
\label{eq:likelihood1}
\end{equation}
where $\bar{C}^{\bf i}_l$ is the source-subtracted cross-power spectrum for
DA pair ${\bf i}$, obtained above, $w^{\bf i}_l$ is the window function for 
spectrum ${\bf i}$, $C^{\rm th}_l$ is now a fixed CMB model spectrum, and 
$\delta A$ is the residual source amplitude.

We can marginalize the likelihood function over the residual point source 
amplitude as follows. Expand equation~(\ref{eq:likelihood1}) as
\begin{eqnarray}
-2\ln{\cal L} & = & \sum_{{\bf ij}\,\,ll'} \,\,
(\bar{C}^{\bf i}_l - C^{\rm th}_lw^{\bf i}_l\,) (\Sigma^{-1})^{\bf ij}_{ll'} (\bar{C}^{\bf j}_{l'}-C^{\rm th}_{l'}w^{\bf j}_{l'}\,) \nonumber \\
 & - & 2(\delta A) \sum_{{\bf ij}\,\,ll'} \,\,
(\bar{C}^{\bf i}_l - C^{\rm th}_lw^{\bf i}_l\,) (\Sigma^{-1})^{\bf ij}_{ll'} S^{\bf j}w^{\bf j}_{l'}\, \nonumber \\
 & + & (\delta A)^2 \sum_{{\bf ij}\,\,ll'} \,\, 
S^{\bf i}w^{\bf i}_l\,(\Sigma^{-1})^{\bf ij}_{ll'} S^{\bf j}w^{\bf j}_{l'}\,,
\end{eqnarray}
which is a quadratic form in $\delta A$, 
${\cal L} \propto \exp[-\case{1}{2}(a(\delta A)^2 - 2b(\delta A) + c)]$, with
\begin{eqnarray}
a & \equiv & \sum_{{\bf ij}\,\,ll'} \,\,
  S^{\bf i}w^{\bf i}_l\,(\Sigma^{-1})^{\bf ij}_{ll'} S^{\bf j}w^{\bf j}_{l'}\,, \nonumber \\
b & \equiv & \sum_{{\bf ij}\,\,ll'} \,\,
  (\bar{C}^{\bf i}_l - C^{\rm th}_lw^{\bf i}_l\,) (\Sigma^{-1})^{\bf ij}_{ll'} S^{\bf j}w^{\bf j}_{l'}\,, \nonumber \\
c & \equiv & \sum_{{\bf ij}\,\,ll'} \,\,
  (\bar{C}^{\bf i}_l - C^{\rm th}_lw^{\bf i}_l\,) (\Sigma^{-1})^{\bf ij}_{ll'} (\bar{C}^{\bf j}_{l'}-C^{\rm th}_{l'}w^{\bf j}_{l'}\,).
\end{eqnarray}
We wish to evaluate the marginalized likelihood, ${\cal L}_A = \int{\cal L}\, 
d(\delta A)$. This is readily evaluated using the substitution $\delta A \to
\delta A - b/a$, giving ${\cal L}_A \propto \exp[-\case{1}{2}(c-b^2/a)]$, where
we drop a  multiplicative term proportional to $a$, which is independent of
$\bar{C}^{\bf i}_l$.  The marginalized likelihood function is thus
\begin{eqnarray}
-2\ln{\cal L}_A & = & \sum_{{\bf ij}\,\,ll'} \,\,
(\bar{C}^{\bf i}_l - C^{\rm th}_lw^{\bf i}_l\,) (\Sigma^{-1})^{\bf ij}_{ll'} (\bar{C}^{\bf j}_{l'}-C^{\rm th}_{l'}w^{\bf j}_{l'}\,) \\
 & - & \frac{1}{a} \sum_{{\bf ij}\,\,ll'} \,\,
(\bar{C}^{\bf i}_l - C^{\rm th}_lw^{\bf i}_l\,) (\Sigma^{-1})^{\bf ij}_{ll'} S^{\bf j}w^{\bf j}_{l'}\,
                   \sum_{{\bf i'j'}\,\,l''l'''} \,\,
(\bar{C}^{\bf i'}_{l''} - C^{\rm th}_{l''}w^{\bf i'}_{l''}\,) (\Sigma^{-1})^{\bf i'j'}_{l''l'''} S^{\bf j'}w^{\bf j'}_{l'''}\,.
\end{eqnarray}
This expression may be recast in the form
\begin{equation}
-2\ln{\cal L}_A = \sum_{{\bf ij}\,\,ll'} \,\, 
(\bar{C}^{\bf i}_l - C^{\rm th}_lw^{\bf i}_l\,)
\,\,(F^{\rm src})^{\bf ij}_{ll'}\,\,
(\bar{C}^{\bf j}_{l'} - C^{\rm th}_{l'}w^{\bf j}_{l'}\,),
\label{eq:likelihood2}
\end{equation}
where $(F^{\rm src})^{\bf ij}_{ll'}$ is
\begin{equation}
(F^{\rm src})^{\bf ij}_{ll'} \equiv (\Sigma^{-1})^{\bf ij}_{ll'}-\frac{1}{a}B^{\bf ij}_{ll'}\;,
\label{eq:Fsrc}
\end{equation}
with
\begin{equation}
B^{\bf ij}_{ll'} \equiv \sum_{{\bf i'j'}\,\,l''l'''}\,\,
(\Sigma^{-1})^{\bf ii'}_{ll''}\,S^{\bf i'}\,w^{\bf i'}_{l''}\,
(\Sigma^{-1})^{\bf jj'}_{l'l'''}\,S^{\bf j'}w^{\bf j'}_{l'''}\,.
\end{equation}
The superscript ``src'' indicates that the Fisher matrix so obtained includes 
point source subtraction uncertainty, in addition to whatever effects have been
included in $(\Sigma^{-1})^{\bf ij}_{ll'}$ already, in this case only cosmic 
variance and noise.  Note that equation~(\ref{eq:Fsrc}) neglects a term
proportional to $\ln\det\,a$ which has a weak dependence on cosmological
parameters.
In the actual calculation, as in the previous section, we 
assume the diagonal form of the covariance matrix in 
equation~(\ref{eq:cl_cov_diag}). 

\section{OPTIMAL WEIGHTING OF MULTI-CHANNEL SPECTRA}
\label{sec:power_opt}

We use the 8 high-frequency differencing assemblies Q1 through W4 to generate
the final combined spectrum. In this Appendix we show how we combine these
spectra into a single estimate of the angular power spectrum.  

The ultimate goal of the \map\ analysis is to produce a likelihood function for
a set of cosmological parameters, $\{\vec{\alpha}\}$, given the data,  
$C^{\bf i}_l$.  Specifically
\begin{equation}
{\cal P}(\vec{\alpha} \vert C^{\bf i}_l) = {\cal L}(C^{\bf i}_l \vert 
C^{\rm th}_l(\vec{\alpha})) {\cal P}(\vec{\alpha}),
\end{equation}
where ${\cal P}(\vec{\alpha} \vert C^{\bf i}_l)$ is the probability of 
$\{\vec{\alpha}\}$ given the data, ${\cal L}(C^{\bf i}_l \vert 
C^{\rm th}_l(\vec{\alpha}))$ is the likelihood of the data given the model,
$C^{\rm th}_l(\vec{\alpha})$, and ${\cal P}(\vec{\alpha})$ is the prior 
probability of the parameter set \citep{verde/etal:2003}.  To this end, we seek
a combined spectrum, $\widehat{C}_l$, that estimates the power spectrum in our
sky, $C^{\rm sky}_l$, with the property that 
$ {\cal P}(\vec{\alpha} \vert \widehat{C}_l) 
= {\cal P}(\vec{\alpha} \vert C^{\bf i}_l)$, and hence
$ {\cal L}(\widehat{C}_l \vert C^{\rm th}_l(\vec{\alpha}))
= {\cal L}(C^{\bf i}_l \vert C^{\rm th}_l(\vec{\alpha}))$.

To estimate the combined spectrum, we approximate the likelihood function for
the cross-power spectra as Gaussian
\begin{equation}
-2\ln{\cal L}(\bar{C}^{\bf i}_l \vert C^{\rm th}_l)
  = \sum_{{\bf ij}\,\,ll'} (\bar{C}^{\bf i}_l - C^{\rm th}_lw^{\bf i}_l)
     \,\,(\Sigma_{\rm full}^{-1})^{\bf ij}_{ll'}\,\,
    (\bar{C}^{\bf j}_{l'} - C^{\rm th}_{l'}w^{\bf j}_{l'}),
\label{eq:likelihoodinit}
\end{equation}
where $\bar{C}^{\bf i}_l$ is the spectrum with the best-fit source model 
subtracted, defined after equation~(\ref{eq:cov_ci_src}), $w^{\bf i}_l$ is the
window function of spectrum ${\bf i}$, defined after
equation~(\ref{eq:cl_obs}), and $\Sigma_{\rm full}$ is
the covariance matrix of the 28 cross-power spectra.  Note that the
treatment in the remainder of this section does not depend on any specific
property of the covariance, so we use generic notation for
readability.  However,  when we generate the \map\ first-year combined
spectrum, the actual form of the  covariance used at this step is 
$(\tilde{\Sigma}_{\rm full})^{\bf ij}_{ll'}$,  where $\tilde{\Sigma}$ indicates the
approximate covariance, obtained in \S\ref{sec:covariance},
that has not yet had the effects of the  foreground mask accounted for.

We seek a spectrum $\widehat{C}_l$ such that
\begin{equation}
-2\ln{\cal L}(\widehat{C}_l \vert C^{\rm th}_l) \equiv \sum_{ll'}
(\widehat{C}_l-C_l^{\rm th})\,Q_{ll'}\,(\widehat{C}_{l'}-C^{\rm th}_{l'})
 = -2\ln{\cal L}(\bar C^{\bf i}_l \vert C^{\rm th}_l),
\label{eq:likelihoodfinal}
\end{equation}
where $Q_{ll'}$ is the inverse-covariance 
matrix of the combined spectrum which comes with the estimate of
$\widehat{C}_l$. Suppose, for simplicity, that $\Sigma_{\rm full}$ is diagonal, 
$(\Sigma_{\rm full})^{\bf ij}_{ll'} = (\Sigma_{\rm full})^{\bf ij}_l \delta_{ll'}$, 
then it is straightforward to show that the deconvolved, weighted-average 
spectrum
\begin{equation}
\widehat{C}_l = \frac{\sum_{\bf ij}\bar{C}^{\bf i}_l
                          (\Sigma_{\rm full}^{-1})^{\bf ij}_lw^{\bf j}_l}
                          {Q_l},
\label{eq:cl_comb_diag}
\end{equation}
with
\begin{equation}
Q_l = \sum_{\bf ij}w_l^{\bf i}(\Sigma^{-1}_{\rm full})^{\bf ij}_lw^{\bf j}_l ,
\label{eq:qll_comb_diag}
\end{equation}
is the desired spectrum.  Substituting equations~(\ref{eq:cl_comb_diag}) and 
(\ref{eq:qll_comb_diag}) into equation~(\ref{eq:likelihoodfinal}) produces
equation~(\ref{eq:likelihoodinit}) up to a term which has a weak dependence
on comological model, which we ignore. This combined spectrum is equivalent 
to the result we would obtain using the ``optimal data compression'' of
\citet{tegmark/taylor/heavens:1997}.

If the inverse covariance matrix is not diagonal in $l$, it can be shown that
the optimal combined spectrum is given by
\begin{equation}
\widehat{C}_l = \frac{\sum_{{\bf ij}\,\,l'}
  \bar{C}^{\bf i}_l\,(\Sigma^{-1}_{\rm full})^{\bf ij}_{ll'}\,C^{\rm fid}_{l'}w^{\bf j}_{l'}}
  {\sum_{l'}Q_{ll'}\,C^{\rm fid}_{l'}},
\label{eq:cl_comb}
\end{equation}
with
\begin{equation}
Q_{ll'} = \sum_{\bf ij}w_l^{\bf i} (\Sigma^{-1}_{\rm full})^{\bf ij}_{ll'}w_{l'}^{\bf j}, 
\label{eq:qll_comb}
\end{equation}
and where $C^{\rm fid}_l$ is a fiducial cosmological model which we take to be
a flat $\Lambda$CDM model with $\Omega_b h^2=0.021$, $\Omega_c h^2=0.129$, 
$\Omega_{\rm tot} = 1$, $h=0.68$, $n_s=1.2$, and $\tau=0.2$.  While this model
has parameters values that are substantially different than the best-fit \map\
model obtained (afterwards) by \citet{spergel/etal:2003}, the parameter 
degeneracies are such that $C^{\rm fid}_l$ is close to the best-fit model for
$l>100$ where this estimator is actually used (see \S\ref{sec:opt_combined}).
The combined spectrum is optimal if the fiducial model chosen is the correct
one; otherwise it is still unbiased but slightly sub-optimal 
\citep{gupta/heavens:2002}.

\section{CUT-SKY FISHER MATRIX}
\label{sec:fisher}

The \map\ sky maps have nearly diagonal pixel-pixel noise covariance
\citep{hinshaw/etal:2003b} which greatly simplifies the properties of the power
spectrum Fisher matrix.  In this Appendix we present an analytic derivation of
the effect of a sky cut and non-uniform pixel weighting.  In
\ref{sec:fisher_analytic}, we assume that we have an optimal estimator of the
power spectrum.  In the noise dominated limit, we can obtain an exact
expression, while in the signal dominated limit, we need to approximate the
signal correlation matrix to obtain an analytic expression.  In
\ref{sec:fisher_interpol}, we interpolate the Fisher matrix between the signal
and noise-dominated regimes and show that it agrees with numerical estimates. 
In \ref{sec:fisher_monte}, we estimate the power spectrum covariance matrix
from Monte Carlo simulations of the sky and calibrate the interpolation
formula.

\subsection{Cut-Sky Fisher Matrix: Analytic Evaluation}
\label{sec:fisher_analytic}

We equate the Fisher matrix of the power spectrum to the curvature of the
likelihood function, equation~(\ref{eq:fisher}), then develop an approximate 
form for the covariance matrix, ${\bf C} = {\bf S+N}$, where ${\bf S}$ is the
signal  matrix and ${\bf N}$ is the noise matrix.  We split the noise matrix
into two pieces: a weight term and a mask term.  In the limit that the pixel
noise is diagonal, the weight term has the form
\begin{equation}
\left({\bf N_w^{-1}}\right)_{ij} = \frac{n_i}{\sigma_0^2} \delta_{ij}
\equiv w_i \delta_{ij}
\end{equation}
where $i$ and $j$ are pixel indices, $n_i$ is the number of observations of 
pixel $i$, $\sigma_0$ is the rms noise of a single observation, and, by 
definition, $w_i$ is the weight of pixel $i$. The mask, $M_i$, is defined so
that $M_i$ equals 0 in pixels that are  not used due to foreground
contamination, and equals 1 otherwise. The noise  matrix can then be written as
the product of the two terms
\begin{equation}
{\bf N^{-1}} = {\bf N_w^{-1}M} = {\bf MN_w^{-1}} 
= \frac{\hat{n}_i}{\sigma_0^2} \delta_{ij} = \hat{w}_i \delta_{ij}
\end{equation}
where  $\hat{n}_i = n_i$ and $\hat{w}_i = w_i$ in the unmasked pixels, and 
are set to 0 otherwise. We thus define the covariance matrix over the full sky, 
which allows us to exploit the orthogonality properties of the spherical 
harmonics. Note that ${\bf M}^2 = {\bf M}$.

The inverse of the full covariance matrix can now be written as
\begin{eqnarray}
{\bf C^{-1}} = {\bf (S+N)^{-1}} & = & {\bf N^{-1}(SN^{-1} + I)^{-1}} \\
 & = & {\bf M^{-1}(SM^{-1} + N_w)^{-1}}.
\end{eqnarray}
The covariance matrix has two limits.  In the noise dominated limit, 
${\bf SN^{-1} \ll I}$,
\begin{equation}
{\bf C^{-1}} \rightarrow {\bf N^{-1}}.
\end{equation}
In the signal dominated limit, ${\bf SM^{-1} \ll N_w}$, only the 
mask alters the covariance matrix, so
\begin{equation}
{\bf C^{-1}} \rightarrow {\bf M S^{-1}M},
\end{equation}
where we have set the inverse of the mask to zero where there is no data, 
i.e. ${\bf M^{-1}} \equiv {\bf M}$.

\subsubsection{Cut-Sky Fisher Matrix in the Noise Dominated Limit}
\label{sec:fisher_noise}

In the noise dominated limit, the Fisher matrix, equation~(\ref{eq:fisher}), is 
approximately
\begin{equation}
F^{\rm N}_{ll'} = \frac{1}{2}{\rm tr}\left({\bf C^{-1}P}_l{\bf C^{-1}P}_{l'}\right)
\rightarrow \frac{1}{2}{\rm tr}\left({\bf N^{-1}P}_l{\bf N^{-1}P}_{l'}\right).
\end{equation}
Using
\begin{equation}
\left({\bf P}_l\right)_{ij} \equiv \frac{\partial{\bf C}}{\partial C_l} 
= \frac{(2l+1)}{4\pi} \, w_l \, P_l(\cos\theta_{ij})
= w_l \, \sum_m Y^*_{lm}(p_i) \, Y_{lm}(p_j)
\label{eq:def_pl_2}
\end{equation}
where $\theta_{ij}$ is the angle between pixels $i$ and $j$, $Y_{lm}(p_i)$ is a
spherical harmonic evaluated in the direction of pixel $i$, and we
have employed the addition theorem for spherical harmonics in the last
equality.  Then the  Fisher matrix takes the form
\begin{equation}
F^{\rm N}_{ll'} = \frac{1}{2}w_l w_{l'} \sum_{mm'} \sum_{ij} 
\hat{w}_i Y^*_{lm}(p_i)Y_{lm}(p_j) \, 
\hat{w}_j Y^*_{l'm'}(p_j)Y_{l'm'}(p_i).
\label{eq:fisher_noise1}
\end{equation}
Now expand the weight array as
\begin{equation}
\hat{w}_i = \sum_{lm} \hat{w}_{lm} \, Y_{lm}(p_i)
\end{equation}
and substitute this into equation~(\ref{eq:fisher_noise1}) to yield
\begin{equation}
F^{\rm N}_{ll'} = \frac{1}{2}w_l w_{l'} \sum_{mm'}\sum_{l''m''}\sum_{l'''m'''} 
\hat{w}_{l''m''}\hat{w}_{l'''m'''} \sum_{ij} 
Y_{l''m''}(p_i) Y^*_{lm}(p_i) Y_{lm}(p_j) \, 
Y_{l'''m'''}(p_j) Y^*_{l'm'}(p_j) Y_{l'm'}(p_i).
\end{equation}
Since $\hat{w}_i$ was defined over the full sky, the sum over pixels may be 
expressed in terms of products of Wigner 3-$j$ symbols.  As shown in 
Appendix~\ref{sec:wigner}, we may use an orthogonality property of the 3-$j$
symbols  to reduce the expression for the Fisher matrix to
\begin{equation}
F^{\rm N}_{ll'} = \frac{1}{2}w_l w_{l'} \, 
\frac{(2l+1)(2l'+1)}{4\pi\Omega^2_p} \sum_{l''}\hat{W}_{l''}\,\wjjj{l}{l'}{l''}{0}{0}{0}^2
\label{eq:fisher_noise}
\end{equation}
where $\hat{W}_l \equiv \sum_m \, \vert\hat{w}_{lm}\vert^2$, and $\Omega_p$ 
is the solid angle per pixel.

\subsubsection{Cut-Sky Fisher Matrix in the Signal Dominated Limit}
\label{sec:fisher_signal}

In the signal dominated limit, the Fisher matrix, equation~(\ref{eq:fisher}), is 
approximately
\begin{equation}
F^{\rm S}_{ll'} = \frac{1}{2}{\rm tr}\left({\bf C^{-1}P}_l{\bf C^{-1}P}_{l'}\right)
\rightarrow \frac{1}{2}{\rm tr}\left({\bf \hat{S}^{-1}P}_l{\bf \hat{S}^{-1}P}_{l'}\right),
\end{equation}
where ${\bf\hat{S}^{-1}} \equiv M_iS^{-1}_{ij}M_j$. Using 
equation~(\ref{eq:def_pl_2}) we can write $F^{\rm S}_{ll'}$ in the form
\begin{equation}
F^{\rm S}_{ll'} = \frac{1}{2}w_l w_{l'} \sum_{mm'}
 \sum_{ijk} M_i M_j S^{-1}_{ij} \,
Y^*_{lm}(p_j) Y_{lm}(p_k)
 \sum_{kp} M_k M_p S^{-1}_{kp} \,
Y^*_{l'm'}(p_p) Y_{l'm'}(p_i).
\label{eq:fisher_signal_1}
\end{equation}
Substituting a normalized expression for ${\bf S^{-1}}$ in the pixel basis, 
\begin{equation}
S^{-1}_{ij} = \Omega^2_p\sum_{lm}\frac{1}{C_l w_l}\,Y^*_{lm}(p_i)Y_{lm}(p_j),
\end{equation}
into equation~(\ref{eq:fisher_signal_1}), one obtains a term
\begin{equation}
  \Omega^2_p \sum_{l''} \frac{1}{C_{l''}w_{l''}}\sum_{m''}
  \sum_i M_i Y_{l'm'}(p_i) Y^*_{l''m''}(p_i)
  \sum_j M_j Y_{l''m''}(p_j) Y^*_{lm}(p_j).
  \label{eq:couplingintegral}
\end{equation}
Since our mask cuts only $\sim$15\% of the sky, the coupling sum,
$\sum_i M_i Y_{l'm'}(p_i)Y^*_{l''m''}(p_i)$, peaks very sharply at 
$|l'-l''|\ll l'$. Therefore, one may approximate 
equation~(\ref{eq:couplingintegral}) with
\begin{eqnarray}
 \nonumber
 & & \frac{\Omega^2_p}{(C_lw_lC_{l'}w_{l'})^{1/2}}\sum_{l''m''}
  \sum_i M_i Y_{l'm'}(p_i) Y^*_{l''m''}(p_i)
  \sum_j M_j Y_{l''m''}(p_j) Y^*_{lm}(p_j) \\
 & = &
  \frac{\Omega_p}{(C_lw_lC_{l'}w_{l'})^{1/2}}  
  \sum_i M_i Y_{l'm'}(p_i) Y^*_{lm}(p_i),
  \label{eq:couplingintegral2}
\end{eqnarray}
where, in the last equality, we have used the completeness relation for 
the spherical harmonics
\begin{equation}
\sum_{lm} Y^*_{lm}(p_i)Y_{lm}(p_j) = \Omega^{-1}_p \delta_{ij}.
\end{equation}
With this approximation equation~(\ref{eq:fisher_signal_1}) reduces to
\begin{equation}
F^{\rm S}_{ll'} \simeq \frac{1}{2} \, \frac{\Omega^2_p}{C_l C_{l'}} \sum_{mm'}
 \sum_{ij} M_i \, Y^*_{lm}(p_i) Y_{lm}(p_j)
        \, M_j \, Y^*_{l'm'}(p_j) Y_{l'm'}(p_i).
\label{eq:fisher_signal_2}
\end{equation}
Following the same steps outlined above for the derivation of
equation~(\ref{eq:fisher_noise}) yields
\begin{equation}
F^{\rm S}_{ll'} = \frac{1}{2} \, \frac{1}{C_l C_{l'}} \, 
\frac{(2l+1)(2l'+1)}{4\pi} \sum_{l''}M_{l''}\,\wjjj{l}{l'}{l''}{0}{0}{0}^2
\label{eq:fisher_signal}
\end{equation}
where $M_l \equiv \sum_m \, \vert m_{lm}\vert^2$, and $m_{lm}$ is the 
spherical harmonic transform of the mask. Note that $M_0=4\pi f^2_{\rm sky}$,
where $f_{\rm sky}$ is the fraction of the sky that survives the mask.

\subsection{Interpolating the Cut-Sky Fisher Matrix}
\label{sec:fisher_interpol}

We can combine the two limiting cases of the Fisher matrix, obtained in the 
previous section, into a single expression
\begin{equation}
F^{\rm mask}_{ll'} = \frac{\sqrt{(2l+1)(2l'+1)}f^2_{\rm sky}}
       {2 (C_l + N_l)(C_{l'}+N_{l'})} \tilde F_{ll'},
\label{eq:coupling}
\end{equation}
where $N_l \equiv \Omega_p \sigma_0^2/(\bar{n}_{obs}w_l)$ is the deconvolved 
noise power spectrum. Using equations~(\ref{eq:fisher_noise}) and 
(\ref{eq:fisher_signal}) derived in the previous section, $\tilde{F}_{ll'}$ 
can be expressed in the low $l$ (signal-dominated) and high $l$ 
(noise-dominated) limits as
\begin{equation}
\tilde F^{\rm S}_{ll'} = \frac{\sqrt{(2l+1)(2l'+1)}}{4\pi f^2_{\rm sky}}
  \sum_{l''} M_{l''}\,\wjjj{l}{l'}{l''}{0}{0}{0}^2,
\label{eq:coupling_signal}
\end{equation}
and
\begin{equation}
\tilde F^{\rm N}_{ll'} = 
 \frac{\sqrt{(2l+1)(2l'+1)}}{4\pi f^2_{\rm sky} \bar{n}^2_{obs}/\sigma_0^4}
 \sum_{l''}\hat{W}_{''}\,\wjjj{l}{l'}{l''}{0}{0}{0}^2.
\label{eq:coupling_noise}
\end{equation}
By construction, these matrices are normalized to 1 on the diagonal, $\tilde
F^{\rm S}_{ll} = \tilde F^{\rm N}_{ll} = 1$, since $M_0 = 4\pi f^2_{\rm sky}$
and  $\hat{W}_0 = 4\pi f^2_{\rm sky} \bar{n}^2_{obs}/\sigma_0^4$.

\begin{figure}
\begin{center}
\includegraphics[angle=90,width=1.0\textwidth]{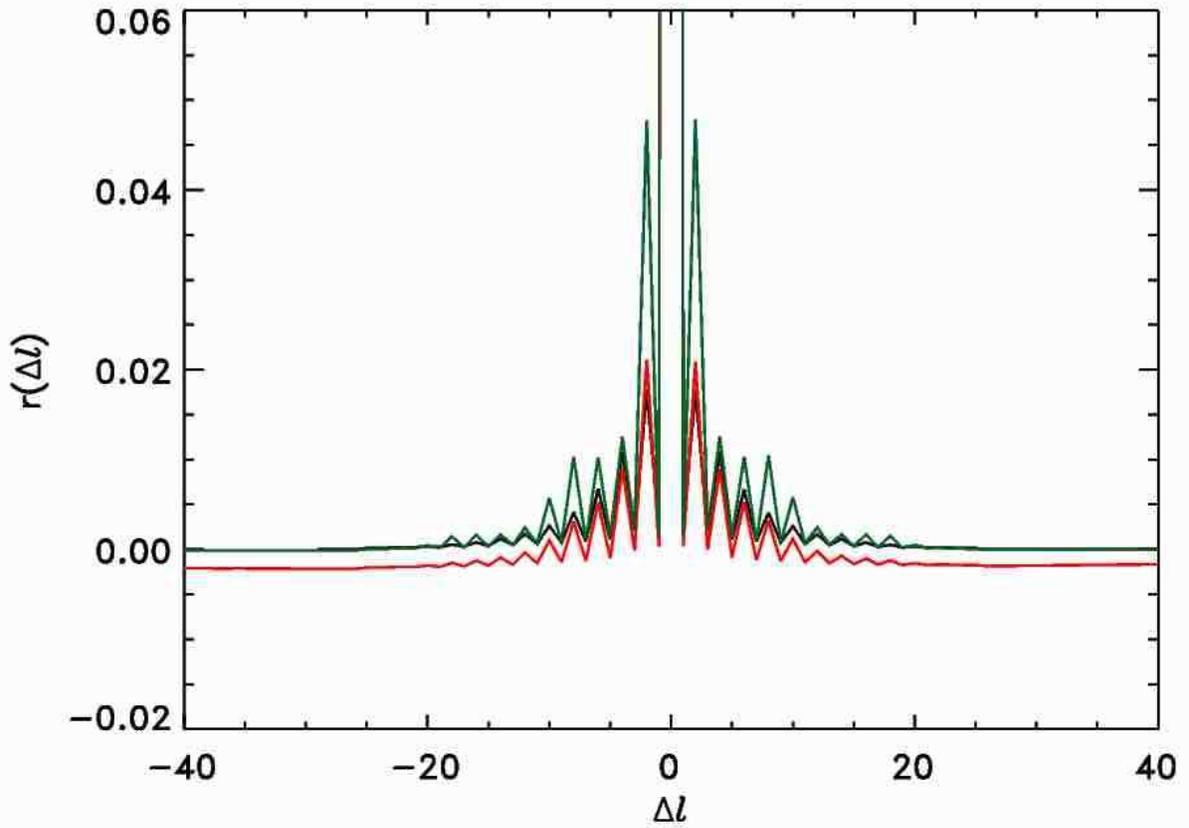}
\end{center}
\caption{Slices of the Fisher (or curvature) matrix normalized as
$F_{ll'}/\sqrt{F_{ll}F_{l'l'}}$, plotted vs.  $\Delta l = l-l'$, from the
Fisher matrix for the combined spectrum.   Black is $l=30$, red is $l=300$,
green is $l=600$.
\label{fig:fisher_slice}}
\end{figure}

We have computed $F^{\rm mask}_{ll'}$ directly, for selected $l$ values, by 
evaluating $Y_{lm}{\bf C}^{-1}Y_{l'm'}$ using the pre-conditioner code
described by \citet{oh/spergel/hinshaw:1999}. We find that deviations between
the numerical and analytical results are consistent with numerical noise in the
Fisher matrix estimate. Figure~\ref{fig:fisher_slice} shows the estimated
Fisher matrix for the Kp2 cut in the noise and the signal dominated limits.  

We interpolate between the two regimes with the following expression:
\begin{equation}
F^{\rm mask}_{ll'} = \frac
  {\left(C_l \sqrt{F^{\rm S}_{l l'}} + N_l \sqrt{F^{\rm N}_{ll'}}\,\right)
  \left(C_{l'} \sqrt{F^{\rm S}_{l l'}} + N_{l'} \sqrt{F^{\rm N}_{ll'}}\,\right)}
  {(C_l + N_l)(C_{l'} + N_{l'})}.
\label{eq:interpolation}
\end{equation} 
Note that the form of the Fisher matrix primarily depends on $\Delta l$,  but
is weakly dependent on $l$ -- there is more coupling in the noise dominated
limit.

\subsection{Power Spectrum Covariance: Monte Carlo Evaluation}
\label{sec:fisher_monte}

As discussed in Appendix~\ref{sec:weights} we compute the $C_l$ using  three
different pixel weightings.  The uniform weighting and the inverse-noise
inverse weighting are optimal in the signal-dominated regime and the 
noise-dominated regime, respectively.  In between these limits the 
transitional weighting performs better. In order to  determine which ranges in
$l$ correspond to which regimes, and to calibrate  our {\em ansatz} for the
covariance matrix, equation~(\ref{eq:coupling}), we  proceed as follows.  Using
100,000 Monte Carlo simulations of signal plus  noise (with \map\ noise levels
and symmetrized beams), we compute the  diagonal elements of the covariance
matrix for the three different weighting  schemes, evaluated with the Kp2 sky
cut.  Denote these estimates  $D^{\rm sim}_l$

We find that the uniform weighting produces the smallest $D^{\rm sim}_l$
below $l=200$.  Inverse-noise weighting is the best above $l=450$, and the 
transitional weighting produces the lowest variance in between.  In each of 
these regimes we use the resulting $D^{\rm sim}_l$ to calibrate our {\em ansatz} 
for the diagonal elements of the covariance matrix
\begin{equation}
D_l = \frac{2}{2l+1}\frac{1}{f_{\rm sky}^{\rm eff}}(C_l + N^{\rm eff}_l)^2,
\end{equation}
as illustrated in \citet{verde/etal:2003}. These calibrations produce a 
smooth correction to equation~(\ref{eq:interpolation}) of at most 6\%.  
No correction at all is required in the signal dominated regime.
 
\section{SOME USEFUL PROPERTIES OF SPHERICAL HARMONICS}
\label{sec:wigner}

The derivation of the form of the Fisher matrix in the signal and noise 
dominated limits led to expressions which included a term
\begin{equation}
\sum_{mm'}\sum_{ij} 
Y_{l''m''}(p_i) Y^*_{lm}(p_i) Y_{lm}(p_j) \, 
Y_{l'''m'''}(p_j) Y^*_{l'm'}(p_j) Y_{l'm'}(p_i),
\label{eq:app_def1}
\end{equation}
where the sum is effectively a double integral over the full sky.  This
can be evaluated in terms of the Wigner 3-$j$ symbols, defined as
\begin{equation}
\int d\Omega_{\n} Y_{lm}(\n) Y^*_{l'm'}(\n) Y_{l''m''}(\n)
\equiv \left(-1\right)^{m'} \sqrt{\frac{(2l+1)(2l'+1)(2l''+1)}{4\pi}}
\wjjj{l}{l'}{l''}{0}{0}{0} \wjjj{l}{l'}{l''}{m}{-m'}{m''}.
\end{equation}
Substituting this into equation~(\ref{eq:app_def1}) gives
\begin{eqnarray}
 \nonumber
 & & \sum_{mm'}\left(-1\right)^{m+m'} \frac{(2l+1)(2l'+1)}{4\pi}\sqrt{(2l''+1)(2l'''+1)}
\wjjj{l}{l'}{l''}{0}{0}{0} \wjjj{l}{l'}{l'''}{0}{0}{0} \\
 & & \times \wjjj{l}{l'}{l''}{m}{-m'}{m''} \wjjj{l}{l'}{l'''}{m}{-m'}{m'''}.
\label{eq:app_3j_2}
\end{eqnarray}
We simplify this using the Wigner 3-$j$ orthogonality condition
\begin{equation}
\sum_{mm'}(-1)^{m+m'}\wjjj{l}{l'}{l''}{-m}{m'}{m''} \wjjj{l}{l'}{l'''}{m}{-m'}{m'''}
 = \frac{1}{2l''+1}\delta_{l''l'''}\delta_{m''m'''}\delta(l,l',l''),
\end{equation}
where $\delta(l,l',l'') = 1$ for $\vert l-l' \vert \le l'' \le l+l'$ and is
0 otherwise.  This reduces equation~(\ref{eq:app_def1}) to
\begin{equation}
\frac{(2l+1)(2l'+1)}{4\pi\Omega^2_p} \wjjj{l}{l'}{l''}{0}{0}{0}^2,
\end{equation}
where the factor of $\Omega^2_p$ accounts for the fact that 
equation~(\ref{eq:app_def1}) is a double sum over pixels, instead of a
double integral.

\clearpage

\clearpage

\clearpage

\clearpage

\clearpage

\clearpage

\clearpage

\clearpage

\clearpage

\clearpage

\clearpage

\clearpage

\clearpage

\clearpage

\clearpage
\begin{deluxetable}{lrrrrrrrr}
\tablecaption{\map\ Power Spectrum Noise Model\tablenotemark{a}}
\tabletypesize{\small}
\tablewidth{0pt}
\tablehead{
\colhead{Parameter} & \colhead{Q1} & \colhead{Q2} & \colhead{V1} & \colhead{V2}
                    & \colhead{W1} & \colhead{W2} & \colhead{W3} & \colhead{W4}}
\startdata
\cutinhead{$l < 200$}
$c_0 (\times 10^2)$ & $-6.604$ & $-6.971$ & $-6.823$ & $-6.674$ & $-7.174$ & $-7.439$ & $-7.044$ & $-7.989$ \\
$c_1 (\times 10^3)$ & $ 1.934$ & $ 4.244$ & $ 2.497$ & $ 1.395$ & $ 3.060$ & $ 3.721$ & $ 1.459$ & $ 4.820$ \\
$c_2 (\times 10^4)$ & $-3.798$ & $-8.521$ & $-5.203$ & $-1.483$ & $-6.021$ & $-6.901$ & $-2.677$ & $-5.870$ \\
$c_3 (\times 10^5)$ & $ 2.549$ & $ 5.921$ & $ 3.715$ & $ 0.242$ & $ 4.147$ & $ 4.540$ & $ 1.750$ & $ 2.167$ \\
\cutinhead{$200 < l < 450$}
$c_0 (\times 10^2)$ & $-6.143$ & $-6.268$ & $-6.320$ & $-6.192$ & $-6.618$ & $-6.833$ & $-6.768$ & $-6.849$ \\
$c_1 (\times 10^4)$ & $-1.334$ & $ 1.341$ & $-0.800$ & $-1.351$ & $ 0.769$ & $ 2.582$ & $ 0.885$ & $ 2.443$ \\
\cutinhead{$l > 450$}
$c_0 (\times 10^2)$ & $-6.202$ & $-6.167$ & $-6.338$ & $-6.264$ & $-6.564$ & $-6.704$ & $-6.688$ & $-6.772$ \\
$c_1 (\times 10^5)$ & $ 3.288$ & $ 4.085$ & $ 3.327$ & $ 5.979$ & $ 7.342$ & $12.970$ & $ 8.851$ & $24.430$ \\
\enddata
\tablenotetext{a}{Best-fit coefficients for the noise model in 
equation~(\ref{eq:noise_fit}).  The units of the output noise are mK$^2$, 
thermodynamic temperature.}
\label{tab:noise_fit}
\end{deluxetable}

\end{document}